%% file: paper.tex
\documentclass[12pt]{article}
\usepackage{epsfig}
\usepackage{amsmath}
\usepackage{hhline}
\usepackage{amssymb}
\usepackage{times}

\newlength{\dinwidth}
\newlength{\dinmargin}
\begin{document}

\newcommand{\Rp}{\mbox{$\not \hspace{-0.15cm} R_p$}}
\newcommand{\lsim}{\raisebox{-1.5mm}{$\:\stackrel{\textstyle{<}}{\textstyle{\sim}}\:$}}
\newcommand{\gsim}{\raisebox{-0.5mm}{$\stackrel{>}{\scriptstyle{\sim}}$}}
\newcommand{\cm}{\mbox{\rm ~cm}}
\def\GeV{\hbox{$\;\hbox{\rm GeV}$}}
\newcommand{\picob}{\mbox{{\rm ~pb}}}
\def\figurename{{\bf Figure}}
\def\tablename{{\bf Table}}
%
%
%
%
%
\pagestyle{empty}
\begin{titlepage}

\begin{flushleft}
DESY 01--021 \hfill ISSN 0418--9833 \\
February 2001
\end{flushleft}

\vspace*{3cm}

\begin{center}
  \Large
  {\bf
    Searches at HERA for Squarks in {\boldmath $R$}-Parity Violating 
    Supersymmetry }

  \vspace*{1cm}
    {\Large H1 Collaboration}
\end{center}

\vspace*{3cm}
 
\begin{abstract}
\noindent
A search for squarks in $R$-parity violating supersymmetry is 
performed in $e^+ p$ collisions at HERA at a centre of mass
energy of $300 \GeV$, using H1  
data corresponding to an integrated luminosity of $37 \picob^{-1}$.
The direct production of single squarks of any generation in 
positron-quark
fusion via a Yukawa coupling $\lambda'$ is considered, taking 
into account {\mbox{$R$-parity}} violating and conserving decays
of the squarks.
No significant deviation from the Standard Model
expectation is found.
The results are interpreted in terms of constraints within
the Minimal Supersymmetric Standard Model (MSSM), the
constrained MSSM and the minimal Supergravity model, and
their sensitivity to the model parameters is studied in detail.
For a Yukawa coupling of 
electromagnetic strength, squark masses
below $260 \GeV$ are excluded at $95 \%$ confidence level
in a large part of the parameter space.
For a 100 times smaller coupling strength masses up to
$182 \GeV$ are excluded. 

\end{abstract}

\vfill

\begin{center}
 {To be submitted to {\em Eur. Phys. J. C}}
\end{center}
\end{titlepage}

\pagestyle{plain}

\include{h1auts_new}
\newpage

\pagestyle{plain}

\section{Introduction}
\label{sec:intro}
 
Supersymmetry (SUSY) relates elementary fermions and bosons and
protects the mass of the Higgs boson from acquiring unnaturally large radiative
corrections. SUSY is often considered an ingredient
of a fundamental theory beyond the Standard Model (SM). It is thus
actively searched for in current experiments.

The $ep$ collider HERA which provides both baryonic and leptonic quantum
numbers in the initial state is ideally suited to search for new particles 
possessing couplings to an electron-quark pair. 
Such particles could be squarks, the scalar SUSY partners of 
quarks, in models where $R$-parity, a discrete symmetry related to
lepton and baryon number conservation,
is violated (\Rp).
These squarks could thus be resonantly produced at HERA via the fusion of 
the initial state positron of energy $27.5 \GeV$ with a quark coming from the 
incident proton of energy $820 \GeV$, up to  
the centre of mass energy $\sqrt{s} \simeq 300 \GeV$.

In this paper, a search is performed for squarks that are singly produced via 
an \Rp\ coupling, considering both \Rp\ decays and decays via
gauge couplings involving neutralinos, charginos or gluinos.
The data  were taken from 1994 to 1997 and
correspond to an integrated luminosity of $37 \picob^{-1}$.
This analysis extends the searches for $eq$ resonances 
previously performed by
H1~\cite{H1LQ99} using the same data sample by
considering specific squark decay modes,
and supersedes earlier published dedicated squark 
searches~\cite{H1RPV96,H1RPVEMINUS} 
which were based on $\sim$ 13 times less data.

\section{Phenomenology}
\label{sec:pheno}

The most general SUSY theory which preserves the gauge invariance 
of the Standard Model allows for Yukawa couplings between 
two known SM fermions and the scalar SUSY partner of a quark
(a squark $\tilde{q}$) or of a lepton
(a slepton $\tilde{l}$).
Such couplings induce violation of the $R$-parity defined as 
$R_p\,=\,(-1)^{3B+L+2S}$, where
$S$ denotes the spin, $B$ the baryon 
number and $L$ the lepton number of the particles. 
Hence $R_p$ is equal to 1 for particles and equal to $-1$ for sparticles.
We consider here the SUSY phenomenology at HERA in the presence of $\Rp$ 
Yukawa couplings but maintain otherwise the minimal field content of 
the  Minimal Supersymmetric Standard Model (MSSM)~\cite{MSSM}.
Of special interest for HERA are the Yukawa couplings
between a squark and a lepton-quark pair~\cite{RPVIOLATION}.
These are described in the
superpotential by the terms $\lambda'_{ijk} L_{i}Q_{j}\bar{D}_k$,
with $i,j,k$ being generation 
indices\footnote{In the usual superfield notation, 
                 $L_{i}$, $Q_{j}$ and $D_k$
                 contain respectively the left-handed leptons,
                 the left-handed quarks 
                 and the right-handed down quark,
                 together with their SUSY partners
                 $\tilde{l}^i_L$, $\tilde{q}^j_L$ and $\tilde{d}^k_R$.}.
The corresponding part of the Lagrangian, expanded in fields, 
reads as:
\begin{eqnarray}
{\cal{L}}_{L_{i}Q_{j}\bar{D}_{k}} &=
   & \lambda^{\prime}_{ijk}
              \left[ -\tilde{e}_{L}^{i} u^j_L \bar{d}_R^k
              - e^i_L \tilde{u}^j_L \bar{d}^k_R - (\bar{e}_L^i)^c u^j_L
     \tilde{d}^{k*}_R \right.           \nonumber \\
 \mbox{} &\mbox{}
 & \left. + \tilde{\nu}^i_L d^j_L \bar{d}^k_R + \nu_L \tilde{d}^j_L
    \bar{d}^k_R + (\bar{\nu}^i_L)^c d^j_L \tilde{d}^{k*}_R \right]
   +\mbox{c.c.}             
 \label{eq:lagrangian}
\end{eqnarray}
where the superscripts $c$ denote the charge conjugate spinors
and the $*$ the complex conjugate of scalar fields.
Hence the couplings
$\lambda'_{1jk}$ allow for resonant production of squarks at HERA through
$eq$ fusion.
For the nine possible $\lambda'_{1jk}$ couplings, the corresponding
single production processes are given in table~\ref{tab:sqprod}.
%
%
\begin{table*}[htb]
  \renewcommand{\doublerulesep}{0.4pt}
  \renewcommand{\arraystretch}{1.2}
 \begin{center}
 \begin{tabular}{p{0.40\textwidth}p{0.60\textwidth}}
         \caption
         {\small \label{tab:sqprod}
         The two resonant squark production processes at HERA ($e^+$ beam)
         allowed by each $R$-parity violating coupling $\lambda'_{1jk}$.} &
   \begin{tabular}{||c||c|c||}
   \hline \hline
   $\lambda'_{1jk}$ & \multicolumn{2}{c||}{production process} \\
   \hline
       & & \\
   111 & $e^+ +\bar{u} \rightarrow \overline{\tilde{d}_R}$
       &$e^+ +d \rightarrow \tilde{u}_L $\\
   112 & $e^+ +\bar{u} \rightarrow \overline{\tilde{s}_R}$
       &$e^+ +s \rightarrow \tilde{u}_L $\\
   113 & $e^+ +\bar{u} \rightarrow \overline{\tilde{b}_R}$
       &$e^+ +b \rightarrow \tilde{u}_L $\\
   121 & $e^+ +\bar{c} \rightarrow \overline{\tilde{d}_R}$
       &$e^+ +d \rightarrow \tilde{c}_L $\\
   122 & $e^+ +\bar{c} \rightarrow \overline{\tilde{s}_R}$
       &$e^+ +s \rightarrow \tilde{c}_L $\\
   123 & $e^+ +\bar{c} \rightarrow \overline{\tilde{b}_R}$
       &$e^+ +b \rightarrow \tilde{c}_L $\\
   131 & $e^+ +\bar{t} \rightarrow \overline{\tilde{d}_R}$
       &$e^+ +d \rightarrow \tilde{t}_L $\\
   132 & $e^+ +\bar{t} \rightarrow \overline{\tilde{s}_R}$
       &$e^+ +s \rightarrow \tilde{t}_L $\\
   133 & $e^+ +\bar{t} \rightarrow \overline{\tilde{b}_R}$
       &$e^+ +b \rightarrow \tilde{t}_L $\\
   \hline \hline
  \end{tabular}
  \end{tabular}
\end{center}
\end{table*}
%
%
%
With an $e^+$ beam, HERA is most sensitive to couplings
$\lambda'_{1j1}$, where mainly $\tilde{u}^j_L$ squarks are being produced
with a cross-section approximately scaling as
{\mbox{$\lambda^{'2}_{1j1} \cdot d(x)$}} where
$d(x)$ is the probability to find a $d$ quark 
in the proton with a momentum fraction
$x=M^2_{\tilde{q}}/s$ and $M_{\tilde{q}}$ denotes the squark mass.
The production of the antisquark 
$\overline{\tilde{d}^k_R}$ 
is also possible
albeit with a much lower cross-section since a
$\bar{u}^j$ antiquark must participate in the fusion.

The search presented here is performed under the simplifying
assumption that one of the $\lambda'_{1jk}$ dominates.
The squarks decay either via their Yukawa coupling into SM fermions (\Rp),
or via their usual gauge couplings (gauge decay) into a 
gluino $\tilde{g}$ (the SUSY partner of the gluon), 
a neutralino
$\chi_{\alpha}^0$ ($\alpha=1,4$) or a 
chargino $\chi_{\beta}^{\pm}$ ($\beta=1,2$).
The mass eigenstates $\chi_{\alpha}^0$ are mixed states of the
photino, the zino and the neutral higgsinos, which are the
SUSY partners of the photon, of the $Z$ and of the two neutral
Higgs fields respectively.
The charginos $\chi_{\beta}^{\pm}$ are mixed states of the charged
higgsinos and of the winos, SUSY partners of the $W^{\pm}$.
Neutralinos, charginos and gluinos are unstable.
This holds in \Rp\ SUSY also for the 
Lightest Supersymmetric Particle (LSP), assumed here to be
a $\chi$ ($\chi^0$ or $\chi^{\pm}$) or a $\tilde{g}$, which decays 
into a quark, an antiquark and a 
lepton~\cite{RPVIOLATION}, via a virtual squark or slepton
undergoing a \Rp\ decay through the $\lambda'$ coupling.
This is in contrast 
to $R_p$ conserving SUSY models
and has important phenomenological consequences.

%
%
 \begin{figure}[htb]
   \begin{center}
    \begin{tabular}{cc}
      \mbox{\epsfxsize=0.45\textwidth 
        \epsffile{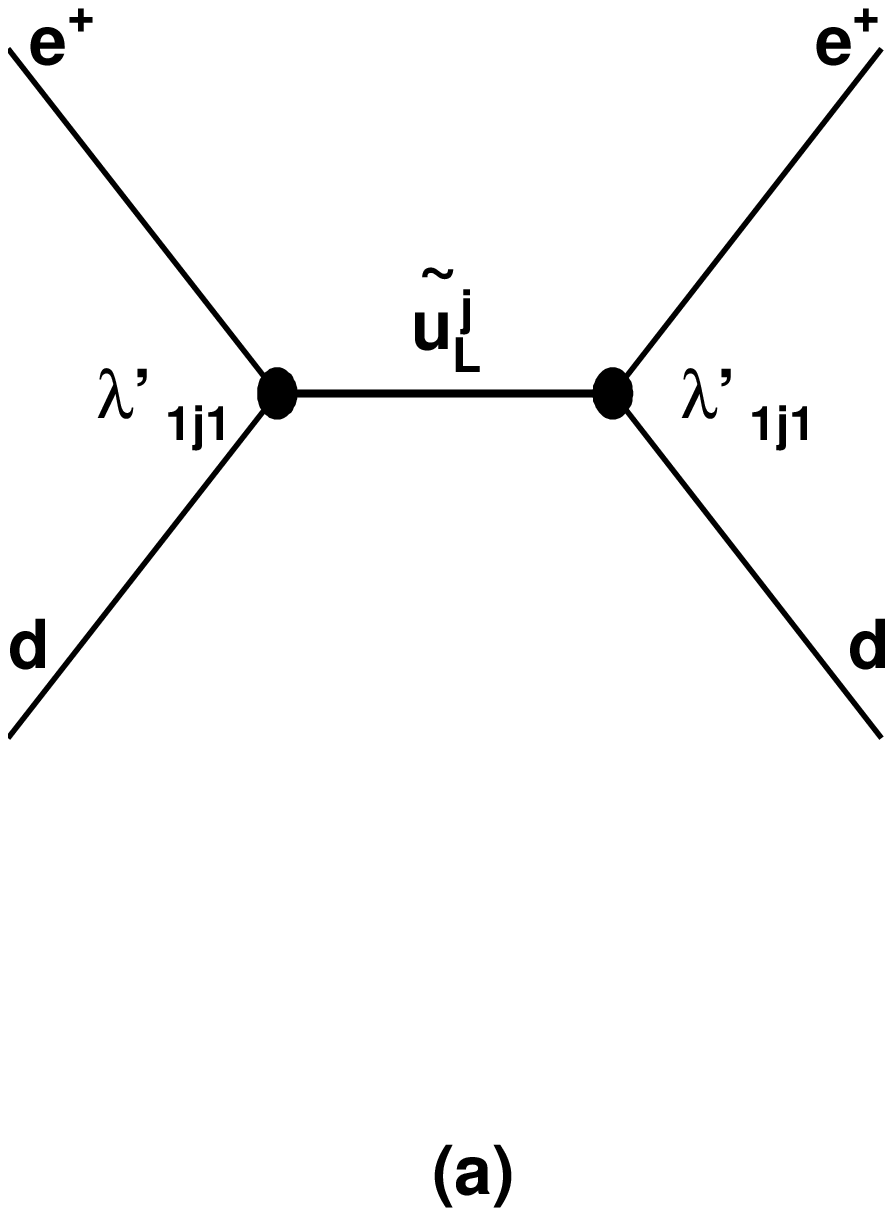}}
    &
      \mbox{\epsfxsize=0.45\textwidth
        \epsffile{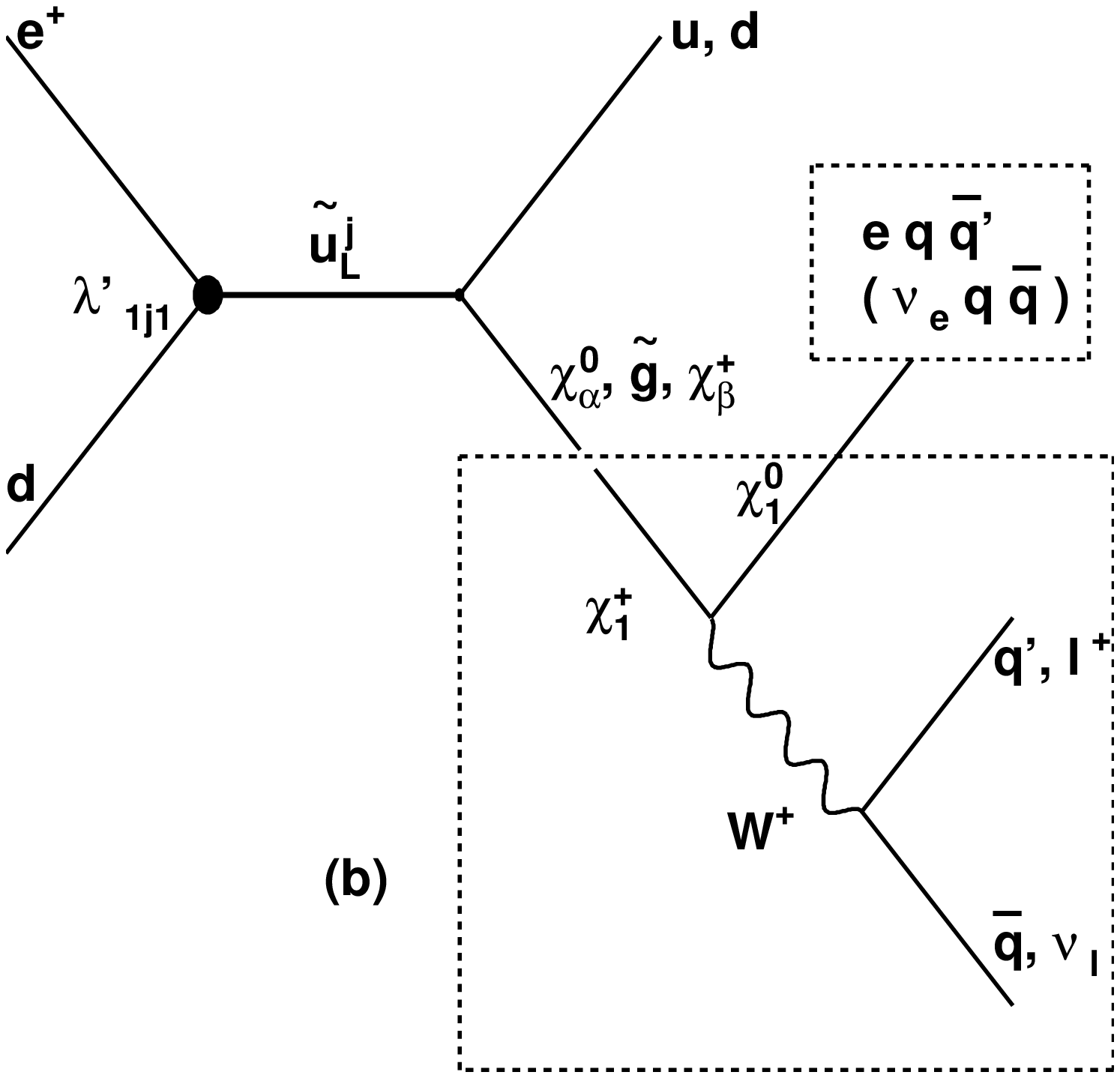}} \\
    &  \\
      \mbox{\epsfxsize=0.45\textwidth
        \epsffile{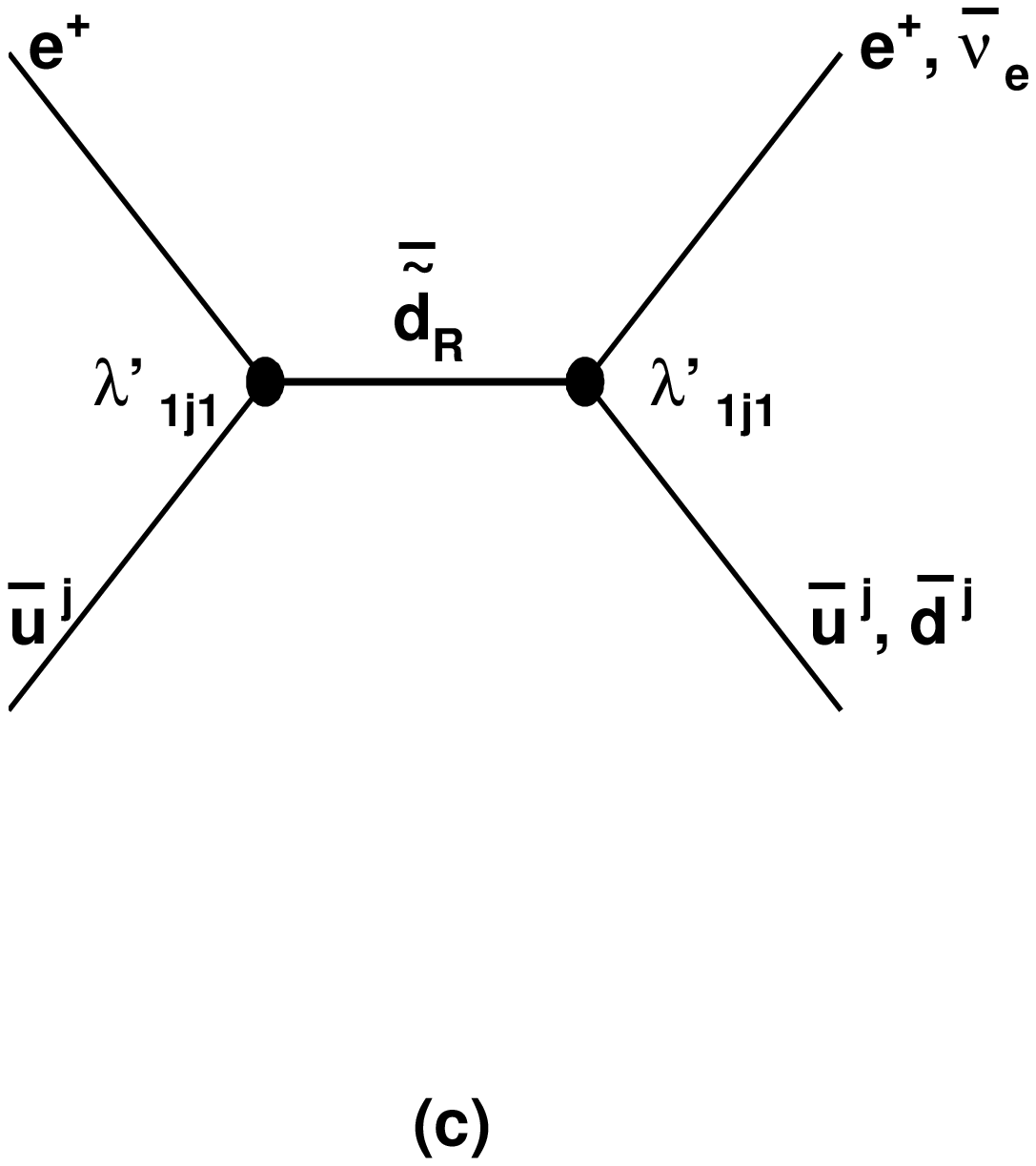}}
    &
      \mbox{\epsfxsize=0.45\textwidth
         \epsffile{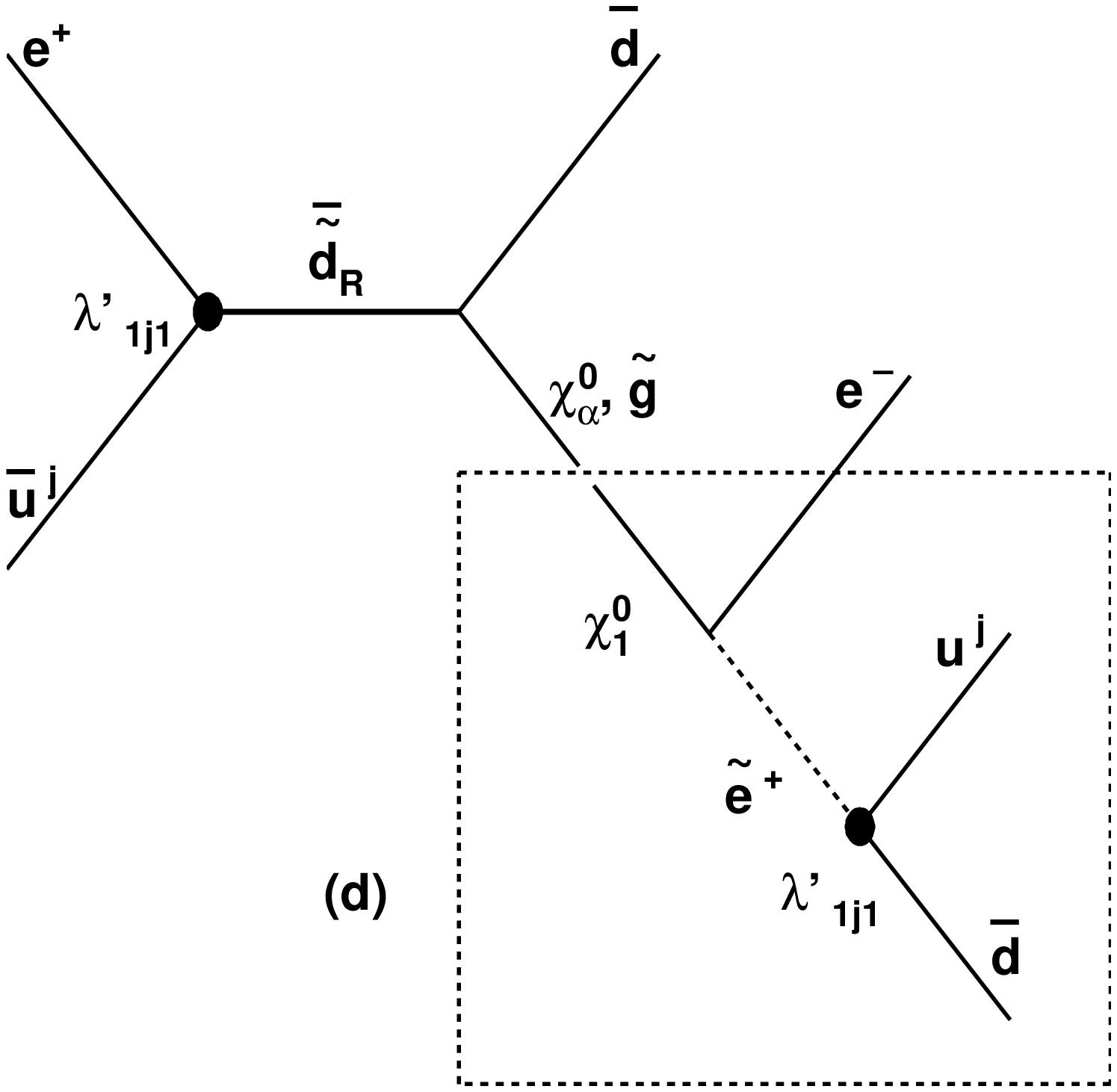}} \\
      \end{tabular}
     \end{center}
  \caption[]{ \label{fig:sqdiag}
     {\small Lowest order $s$-channel diagrams for \Rp\
       squark production at HERA followed by
       (a), (c) \Rp\ decays and (b), (d) gauge decays
       of the squark.
       In (b) and (d), the emerging neutralino, chargino 
       or gluino might
       subsequently undergo a $R_p$ violating or
       $R_p$ conserving decay of which examples are
       shown in the dashed boxes for (b) the $\chi_1^{+}$ and
       (d) the $\chi_1^0$. }}
 \end{figure}
%

 
In cases where both production and decay occur through a
$\lambda'_{1jk}$ coupling (e.g.\ Fig.~\ref{fig:sqdiag}a and c for
$\lambda'_{1j1} \ne 0$), the squarks have the same signature as scalar
leptoquarks (LQ)~\cite{BUCHMULL}.
As can be seen from equation~(\ref{eq:lagrangian}),
the $\overline{\tilde{d}^k_R}$ can decay either into $e^+ + {\bar{u}}^j$ or
$\bar{\nu}_e + {\bar{d}}^j$, while the $\tilde{u}^j_L$ only decays into
$e^+ + d^{k}$.
The final state signatures consist of a lepton and a jet and
are, event-by-event, indistinguishable from 
SM neutral current (NC) and charged current (CC)
Deep-Inelastic Scattering (DIS).

 
When the $\tilde{u}^j_L$ ($\overline{\tilde{d}^k_R}$)
undergoes
a gauge decay into a $\chi^0$, a $\chi^+$ or
a $\tilde{g}$
(a $\chi^0$ or a $\tilde{g}$)
as shown in Fig.~\ref{fig:sqdiag}b and d,
the final state will depend on their subsequent decays.
Neutralinos $\chi^0_{\alpha}$ with $\alpha > 1$ as well as
charginos (gluinos) usually undergo 
gauge decays into a lighter $\chi$ and two SM fermions (two quarks),
through a real or virtual boson or sfermion (squark).
The decay chain ends with the \Rp\ decay of one sparticle,
usually that of the LSP.

\Rp\ decays of $\chi$'s or 
gluinos are mainly relevant
for the lightest states.
Neutralinos can undergo the \Rp\ decays
$\chi^0 \rightarrow e^{\pm} q \bar{q}'$ or
$\chi^0 \rightarrow \nu q \bar{q}$, 
the former (latter) being more frequent if the $\chi^0$ is
dominated by its photino (zino) component.
Gluinos can undergo the same \Rp\ decays.
When a $\chi^0$ or a $\tilde{g}$
decays via \Rp\ into a charged lepton, both the
``right'' and the ``wrong'' charge lepton (with respect to the incident
beam) are equally probable, this latter case leading to largely background free
striking signatures for lepton number violation.
In contrast, the only possible \Rp\ decays for charginos 
are $\chi^+ \rightarrow \bar{\nu} u^k \bar{d}^j$ and
$\chi^+ \rightarrow e^+ d^k \bar{d}^j$.


The decay chains of  $\tilde{u}^j_L$ and
$\overline{\tilde{d}^k_R}$ analysed in this paper
are classified into seven distinguishable event
topologies as described in table~\ref{tab:sqtopo1}.
This classification relies on the number of charged leptons and/or jets
in the final state, and on the presence of missing energy.
Channels labelled {\boldmath{$LQe$}} and {\boldmath{$LQ\nu$}} 
are the ``leptoquark-like" decay
modes of the squark, proceeding directly via \Rp,
while the remaining channels cover the gauge decays of the squark
and are characterised by multijet (MJ) final states.
Channels labelled 
{\boldmath{$eM\!J$}}, {\boldmath{$e^- M\!J$}} and {\boldmath{$\nu M\!J$}}
involve one or two SUSY fermions ($\chi$ or $\tilde{g}$)
denoted by $X$ and $Y$ in  table~\ref{tab:sqtopo1}.
Channels {\boldmath{$e \ell M\!J$}} and {\boldmath{$\nu \ell M\!J$}} 
necessarily involve
two SUSY fermions.
Decay patterns involving more than two
$\chi$ or $\tilde{g}$ are 
kinematically suppressed and are not searched for explicitly.
The relative contributions 
of the channels considered
depend in particular on the value of the Yukawa coupling $\lambda'$ 
and on the gaugino\footnote{The gauginos are the SUSY partners
of the gauge bosons.}-higgsino 
mixture of neutralinos and charginos.
They will be shown as functions of the squark mass in
section~\ref{sec:unconstrained} for some example cases.

Additional event topologies not listed in table~\ref{tab:sqtopo1}
could in principle arise in the case where the $\chi^0_1$ has 
such a small
decay width (e.g.\ when it has large higgsino components)
that it decays far away from the interaction point or leads to final states
with displaced vertices~\cite{H1RPV96}.
However the  region of MSSM parameter space which would allow
a $\chi^0$ to escape detection
for a finite value of the \Rp\ coupling
is now very severely constrained by the searches for charginos
carried out at LEP~\cite{LEPNOS6RPC,L3RPV}.
The lifetimes of the sparticles are neglected
in this analysis.

%
%
%
\begin{table}[htb]
 \renewcommand{\doublerulesep}{0.4pt}
 \renewcommand{\arraystretch}{1.0}
 \begin{center}
  \begin{tabular}{||c|l|c||}
  \hline \hline
  {\bf Channel}  &  \multicolumn{1}{c|}{\bf Decay processes}
           & \multicolumn{1}{c||}{\bf Signature} \\
           & \multicolumn{1}{c|}{ }
           & \multicolumn{1}{c||}{ }         \\ \hline
  {\boldmath{$LQe$}} &  \begin{tabular}{cccccc}
          $\tilde{q}$ & $\stackrel{\lambda'}{\longrightarrow}$
                      & $e^+$   & $q$    &    &
        \end{tabular}
     &  \begin{tabular}{c}
        high $P_T$ $e^+$ + 1 jet
        \end{tabular} \\                                        \hline
  {\boldmath{$LQ\nu$}} &  \begin{tabular}{cccccc}

         $\overline{\tilde{d}^k_R}$ & $\stackrel{\lambda'}{\longrightarrow}$
                      & $\bar{\nu}_e$   & $\bar{d}$     &    & 
        \end{tabular}
     &  \begin{tabular}{c}
         missing $P_T$ + 1 jet
        \end{tabular} \\                                        \hline
  {\boldmath{$eM\!J$}} & \begin{tabular}{ccccll}
         $\tilde{q}$ & $\longrightarrow$
                     & $q$ & $X$  &  &  \\
         &  &        & $\stackrel{\lambda'}{\hookrightarrow}$
                     & $e^+ \bar{q} q$ & \\
         $\tilde{q}$ & $\longrightarrow$
                     & $q$ & $X$ &  & \\
         &  &        & $\hookrightarrow$
                     & $q \bar{q}$ & $\hspace{-0.5cm}Y$ \\
         &  &  &  & 
                  & $\hspace{-0.5cm}
                \stackrel{\lambda'}{\hookrightarrow}$ $e^+ \bar{q} q$  \\
       \end{tabular}
     & \begin{tabular}{c}
         $e^+$ \\
        + multiple jets
       \end{tabular}\\                                          \hline
  {\boldmath{$e^-\!M\!J$}} & \begin{tabular}{ccccll}
         $\tilde{q}$ & $\longrightarrow$
                     & $q$ & $\chi^0_{\alpha}, \tilde{g}$ &  & \\
         &  &        & $\stackrel{\lambda'}{\hookrightarrow}$
         & $e^- \bar{q} q$ &  \\
         $\tilde{q}$ & $\longrightarrow$ & $q$ & $X$ &  & \\
         &  &        & $\hookrightarrow$
                     & $q \bar{q}$ & $\hspace{-0.5cm}Y$ \\
         &  &  & & 
         & $\hspace{-0.5cm}
           \stackrel{\lambda'}{\hookrightarrow}$ $e^- \bar{q} q$ \\
       \end{tabular}
     & \begin{tabular}{c}
         $e^-$ \\ (i.e.\ wrong sign lepton) \\
        + multiple jets
       \end{tabular}\\                                          \hline
  {\boldmath{$\nu M\!J$}} & \begin{tabular}{ccccll}
         $\tilde{q}$ & $\longrightarrow$
                     & $q$ & $X$ &  & \\
         &  &        & $\stackrel{\lambda'}{\hookrightarrow}$
         & $\nu  \bar{q} q$ &  \\
         $\tilde{q}$ & $\longrightarrow$ & $q$ & $X$ &  & \\
         &  &        & $\hookrightarrow$
                     & $q \bar{q}$ & $\hspace{-0.5cm}Y$ \\
         &  &  & &
         & $\hspace{-0.5cm}
           \stackrel{\lambda'}{\hookrightarrow}$ $\nu  \bar{q} q'$ \\
       \end{tabular}
     & \begin{tabular}{c}
        missing $P_T$ \\
        + multiple jets
       \end{tabular}\\                                           \hline
  {\boldmath{$e \ell M\!J$}} & \begin{tabular}{ccccll}
         $\tilde{q}$ & $\longrightarrow$ & $q$ & $X$ &  & \\
         &  &        & $\hookrightarrow$
                     & $\ell \nu_{\ell} $ & $\hspace{0cm}Y$ \\
         &  &  & &
         & $\hspace{0cm}
           \stackrel{\lambda'}{\hookrightarrow}$ $e^{\pm} \bar{q} q$ \\

         $\tilde{q}$ & $\longrightarrow$ & $q$ & $X$ &  & \\
         &  &        & $\hookrightarrow$
                     & $\ell^+ \ell^-  $ & $\hspace{0cm}Y$ \\
         &  &  & &
         & $\hspace{0cm}
           \stackrel{\lambda'}{\hookrightarrow}$ $e^{\pm} \bar{q} q$ \\

         $\tilde{q}$ & $\longrightarrow$ & $q$ & $X$ &  & \\
         &  &        & $\hookrightarrow$
                     & $e^+ e^-  $ & $\hspace{0cm}Y$ \\
         &  &  & &
         & $\hspace{0cm}
           \stackrel{\lambda'}{\hookrightarrow}$ $\nu \bar{q} q$ 
       \end{tabular}
     & \begin{tabular}{c}
         $e$ \\
        +  $\ell$ ($e$ or $\mu$) \\
        + multiple jets
       \end{tabular}\\                                          \hline
  {\boldmath{$\nu \ell M\!J$}} & \begin{tabular}{ccccll}
         $\tilde{q}$ & $\longrightarrow$ & $q$ & $X$ &  & \\
         &  &        & $\hookrightarrow$
                     & $\ell \nu_{\ell} $ & $\hspace{0cm}Y$ \\
         &  &  & &
         & $\hspace{0cm}
           \stackrel{\lambda'}{\hookrightarrow}$ $\nu \bar{q} q$ \\

         $\tilde{q}$ & $\longrightarrow$ & $q$ & $X$ &  & \\
         &  &        & $\hookrightarrow$
                     & $\mu^+ \mu^- $ & $\hspace{0cm}Y$ \\
         &  &  & &
         & $\hspace{0cm}
           \stackrel{\lambda'}{\hookrightarrow}$ $\nu \bar{q} q$
       \end{tabular}
     & \begin{tabular}{c}
          $\ell$ ($e$ or $\mu$) \\
        + missing $P_T$ \\
        + multiple jets
       \end{tabular}\\                                          
   \hline \hline
  \end{tabular}
  \caption[]
          {\small \label{tab:sqtopo1}
               Squark decay channels in \Rp\ SUSY classified by
               distinguishable event topologies. 
               $X$ and $Y$ denote  a neutralino,
               a chargino or a gluino. Quarks are generically denoted
               by $q$, except for the 
               {\boldmath{$LQ \nu$}} channel
               which involves specific (s)quark flavours.
               The final states corresponding to $\ell = \tau$
               for the {\boldmath{$e \ell M\!J$}} and
               {\boldmath{$\nu \ell M\!J$}} channels are
               not explicitly looked for in this analysis. }
 \end{center}
\end{table}

\section{The H1 Detector}
\label{sec:h1det}
 
A detailed description of the H1 detector can be found
in~\cite{H1DETECT}.
Here we describe only the components relevant for the present analysis
in which the final state of the events involves either
a positron\footnote{Unless otherwise stated, the analysis does not distinguish 
                    explicitly between $e^+$ and $e^-$.}
with high transverse energy or a large amount of hadronic transverse
energy flow.

The positron energy and angle are measured in a liquid argon (LAr) sampling
calorimeter~\cite{H1LARCAL} covering the polar angular
range 4$^{\circ} \le \theta \le$ 154$^{\circ}$ and all azimuthal angles.
Polar angles are defined by taking the origin of the coordinate system to be
at the nominal interaction point and the $z$-axis in the direction of the
proton beam.
The granularity of the LAr calorimeter
is optimised to provide fine and approximately uniform
segmentation in laboratory pseudorapidity $\eta$ and azimuthal angle $\phi$.
The calorimeter consists of a lead/argon electromagnetic section
followed by a stainless steel/argon hadronic section.
Test beam measurements~\cite{H1CALRES} of the LAr calorimeter modules
have shown an energy resolution of 
$\sigma(E)/E \simeq$ $12\%$/$\sqrt{E/\GeV} \oplus1\%$ for electrons
and $\sigma(E)/E \simeq$ $50\%$/$\sqrt{E/\GeV} \oplus2\%$
for pions.
The angular resolution on the positron measured from the
electromagnetic shower in the calorimeter varies from $\sim 2$ mrad
below 30$^{\circ}$ to $\lsim 5$ mrad at larger angles.
For the acquisition of events we rely on
the LAr trigger system~\cite{H1LARCAL} whose efficiency is
close to $100 \%$ for the transverse energies ($E_T$) considered here.
A lead/scintillating-fibre backward calorimeter~\cite{H1SPACAL} extends the
coverage\footnote{The detectors in the backward region were upgraded
                  in 1995 by the replacement of the lead/scintillator tile
                  calorimeter~\cite{H1BEMC} and a proportional chamber.}
at larger angles (153$^{\circ} \le \theta \lsim$ 178$^{\circ}$).

The tracking system which is surrounded by the calorimeters 
is used in particular
to determine the position of the interaction vertex.
The main components of this system are central drift and proportional
chambers (25$^{\circ} \le \theta \le$ 155$^{\circ}$), a forward track
detector  (7$^{\circ} \le \theta \le$ 25$^{\circ}$) and a backward
drift chamber.
The tracking chambers and calorimeters are surrounded by a superconducting
solenoid providing a uniform field of $1.15${\hbox{$\;\hbox{\rm T}$}}
parallel to the $z$ axis within the detector volume.
The instrumented iron return yoke surrounding this solenoid is used to
measure leakage of hadronic showers and to recognise muons.
In the very forward region ($\theta \le 15^{\circ}$) 
muons can also be detected in three double layers of drift  chambers,
forming the Forward Muon Detector.
The luminosity is determined from the rate of Bethe-Heitler
$e p \rightarrow e p \gamma$ bremsstrahlung events measured in a luminosity
monitor.

\section{Monte Carlo Event Generation}
\label{sec:dismc}


For each possible SM background source, complete Monte Carlo
simulations of the H1 detector response are performed.
Most of them correspond to a luminosity of more than 10 times 
that of the data.


For the simulation of the NC and CC DIS backgrounds, 
the DJANGO~\cite{DJANGO} event generator is used, which
includes first order QED radiative corrections.
QCD radiation is treated following 
the approach of the Colour Dipole Model~\cite{CDM} 
and is implemented using
ARIADNE~\cite{ARIADNE}.
The hadronic final state is generated using the
string fragmentation model~\cite{JETSET74}.
The parton densities in the proton used to estimate DIS 
expectations are taken
from the MRST~\cite{MRST} parametrisation. 

For  direct and resolved photoproduction ($\gamma p$) 
of light and heavy flavours,
the PYTHIA event generator~\cite{PYTHIA} 
is used which relies on first order QCD matrix elements
and uses leading-log parton showers
and string fragmentation~\cite{JETSET74}.
The GRV  (GRV-G) parton densities~\cite{SFGRVGLO} in the proton
(photon) are used.

 
The simulation of the leptoquark-like signatures ({\boldmath $LQe$} and
{\boldmath $LQ\nu$}) relies on the event generator LEGO~\cite{LEGOSUSS}
which is described in more detail in~\cite{H1RPVEMINUS,H1LQ}. 
For squarks undergoing gauge decays, we use the SUSYGEN~\cite{SUSYGEN} 
event generator, recently extended~\cite{SUSYGEN3} to allow 
the generation of SUSY events in $ep$ collisions.
Any gauge decay of the squark can be generated, and the
cascade decays of the subsequent $\chi$'s or $\tilde{g}$ 
are performed according to the corresponding matrix elements.

In both LEGO and SUSYGEN,
initial and
final state parton showers are simulated following the
DGLAP~\cite{DGLAP} evolution equations, 
and string fragmentation~\cite{PYTHIA,JETSET74}
is used for the non-perturbative part of the hadronisation.
In addition initial state bremsstrahlung in the collinear approximation 
is simulated in the LEGO generator.
The parton densities used~\cite{MRST} are evaluated at the
scale of the squark mass.
This scale is also chosen for the maximum virtuality of parton showers
initiated by a quark coming from the squark decay.
Moreover, in the SUSYGEN generator, the parton showers modelling QCD
radiation off quarks emerging from a $\chi$ or $\tilde{g}$ decay are
started at a scale given by the mass of this sparticle.

To allow a model independent interpretation of the results,
the signal topologies given in table~\ref{tab:sqtopo1} 
were simulated for a wide range
of masses of the SUSY particles.
The events were passed through a complete simulation of the H1 detector.
The squark mass was varied from
$75 \GeV$ to $275 \GeV$ in steps of typically $25 \GeV$. 
Gauge decays of squarks involving one or two SUSY fermions 
($\chi$ or $\tilde{g}$) were
simulated separately.
For gauge decays of squarks into a $\chi^0$, a $\chi^+$ or
a $\tilde{g}$ which directly decays
via \Rp\ (i.e.\ processes corresponding to the first line of the
{\boldmath{$eM\!J$}}, {\boldmath{$e^-\!M\!J$}} and {\boldmath{$\nu M\!J$}} 
rows in
table~\ref{tab:sqtopo1})
the process $\tilde{q} \rightarrow q \chi^0_1$ was
simulated for $\chi^0_1$ masses ranging between $40 \GeV$  and
$160 \GeV$.
In order to study gauge decays involving two $\chi$ or $\tilde{g}$,
the process 
$\tilde{q} \rightarrow q \chi^+_1 \rightarrow q \chi^0_1 f \bar{f'}$
was simulated for $\chi^+_1$ masses ranging between $90 \GeV$
and $\sim M_{\tilde{q}}$, and for 
$\chi^0_1$ masses between half of the $\chi^+_1$ mass
and $\sim M_{\chi^+_1}$. 
Masses of the $\chi$'s were varied in steps of about $20 \GeV$.
These simulations allowed the determination of signal selection efficiencies
as a function of the 
masses of the squark and of the involved $\chi$ or $\tilde{g}$
for essentially all allowed scenarios, since
the grid size chosen for the simulated scenarios was
small enough for a linear interpolation between them.

\section{Event Selection and Comparison with Standard Model \\
         Expectation}
\label{sec:analyz}

The data reduction starts by the rejection of non-$ep$ background, 
which is common to all channels presented below.
It is required
that the events are accepted by a set of beam halo and cosmic muon 
filters~\cite{H1F2PAPER}, that
they satisfy constraints on their timing relative to the 
nominal time of the beam bunch crossings, and
that a primary interaction vertex is reconstructed. 

Events containing lepton(s), hard jets, or a
large amount of missing transverse energy are then
selected using the following identification criteria:
\begin{itemize}
\item
  a {\bf{positron}} (or electron) is 
  identified by a shower shape analysis of clustered energy 
  deposits in the
  LAr calorimeter;
        the positron energy cluster should contain more than $98\%$ of
        the LAr energy found within a pseudorapidity-azimuthal cone
        centered around the positron direction and
        of opening $\sqrt{ (\Delta \eta)^2 + (\Delta \phi)^2 } = 0.25$,
        where
        $\eta = -\ln \tan \frac{\theta}{2}$;
        at least one charged track is required within this isolation
        cone;
\item
a {\bf{muon}} candidate is identified as a track measured  in the   
central  and/or forward tracking system, which 
has to match geometrically an energy deposit
in the LAr calorimeter compatible with  a 
minimum ionising particle, 
and/or a track in the instrumented iron and/or a track in the forward 
muon detector;
\item
 {\bf{hadronic jets}} are reconstructed 
         from energy deposits in the LAr calorimeter
         using a cone algorithm in the laboratory frame
         with a radius $\sqrt{\Delta \eta^2 + \Delta \phi^2} = 1$;
         the fraction of the jet energy deposited in the hadronic part
         of the calorimeter must be at least $5 \%$;
\item
the {\bf{missing transverse momentum}} $P_{T,miss}$ is obtained as
  \begin{eqnarray}
  \label{eq:ptmiss}
     P_{T,miss} \equiv  \sqrt{ 
        \left(\sum E_i \sin \theta_i \cos \phi_i \right)^2
      + \left(\sum E_i \sin \theta_i \sin \phi_i \right)^2 } 
  \end{eqnarray}
    where the summation runs over all energy deposits $i$ in the
    calorimeters.
\end{itemize}

In addition, the selection makes use of the following kinematic
variables:
\begin{itemize}
\item the momentum balance with respect to the direction of the
  incident positron,
      obtained as:
      \begin{eqnarray}
      \label{eq:empz}
      \sum \left(E - P_z\right) \equiv
      \sum E_i \left( 1 - \cos \theta_i \right) 
      \end{eqnarray}
       where the summation runs over all energy deposits $i$ in the
    calorimeters.
      $\sum \left(E - P_z\right)$ should peak at 
      twice the energy $E^0_e$ of the incident positron for events
      where only particles escaping in the proton direction remain undetected;
\item the Lorentz invariants $y$, $Q^2$ and $x$ characterising the kinematics
 of a DIS reaction, as well as the energy $M$ in the centre
 of mass of the hard subprocess, are
 determined using the measurement of the polar angle $\theta_e$,
 the energy $E_e$ 
 and the transverse energy $E_{T,e}$ of the highest
 $E_T$ positron:
  $$ y_e = 1 - \frac{E_e (1 - \cos \theta_e) }{2E_e^0}, \;\;\;\;
     Q^2_e = \frac{E^2_{T,e}}{1-y_e}, \;\;\;\;
     x_e = \frac{Q^2_e} {y_e s}, \;\;\;\;
     M_e = \sqrt{x_e s} \;\; ; \;\;$$
\item the variables $y$, $Q^2$, $x$ and $M$ calculated using the Jacquet-Blondel
 ansatz~\cite{JACQUET}:
 $$ y_h=\frac{\sum \left(E-P_z\right)_h}{2E_e^0}, \;\;\;\;
    Q^2_h= \frac{P^2_{T,h}}{1-y_h},\;\;\;\;
    x_h = \frac{Q^2_h} {y_h s}, \;\;\;\;
    M_h = \sqrt{x_h s} \;\; ; \;\;$$
where $P_{T,h}$ and $\sum \left(E-P_z\right)_h$ are calculated
as in equations~(\ref{eq:ptmiss}) and~(\ref{eq:empz}), 
but restricting the summations to all measured hadronic final
state energy deposits.

\end{itemize}

The search for squarks decaying via \Rp\ couplings into channels
{\boldmath $LQe$} and {\boldmath $LQ\nu$} is identical to
the search for first generation leptoquarks presented in~\cite{H1LQ99}.
Gauge decay channels are grouped into two classes, 
$e + {\mbox{ jets }} + X$ and $\nu + {\mbox{ jets }} + X$.
Preselection criteria are designed for these two classes
of events, on top of
which dedicated cuts are applied for the gauge decay channels
listed in table~\ref{tab:sqtopo1}.
For all considered channels, the selection criteria 
are given in table~\ref{tab:Sicuts} together
with the resulting signal efficiencies and
the numbers of observed and expected events.

\subsection{Analysis of Squark {\boldmath{$R$}}-Parity Violating Decays}
\label{sec:selS1}

\begin{figure}[htb]
  \begin{center}
  \begin{tabular}{cc}
     \hspace*{-0.9cm}\mbox{\epsfxsize=0.55\textwidth
         \epsffile{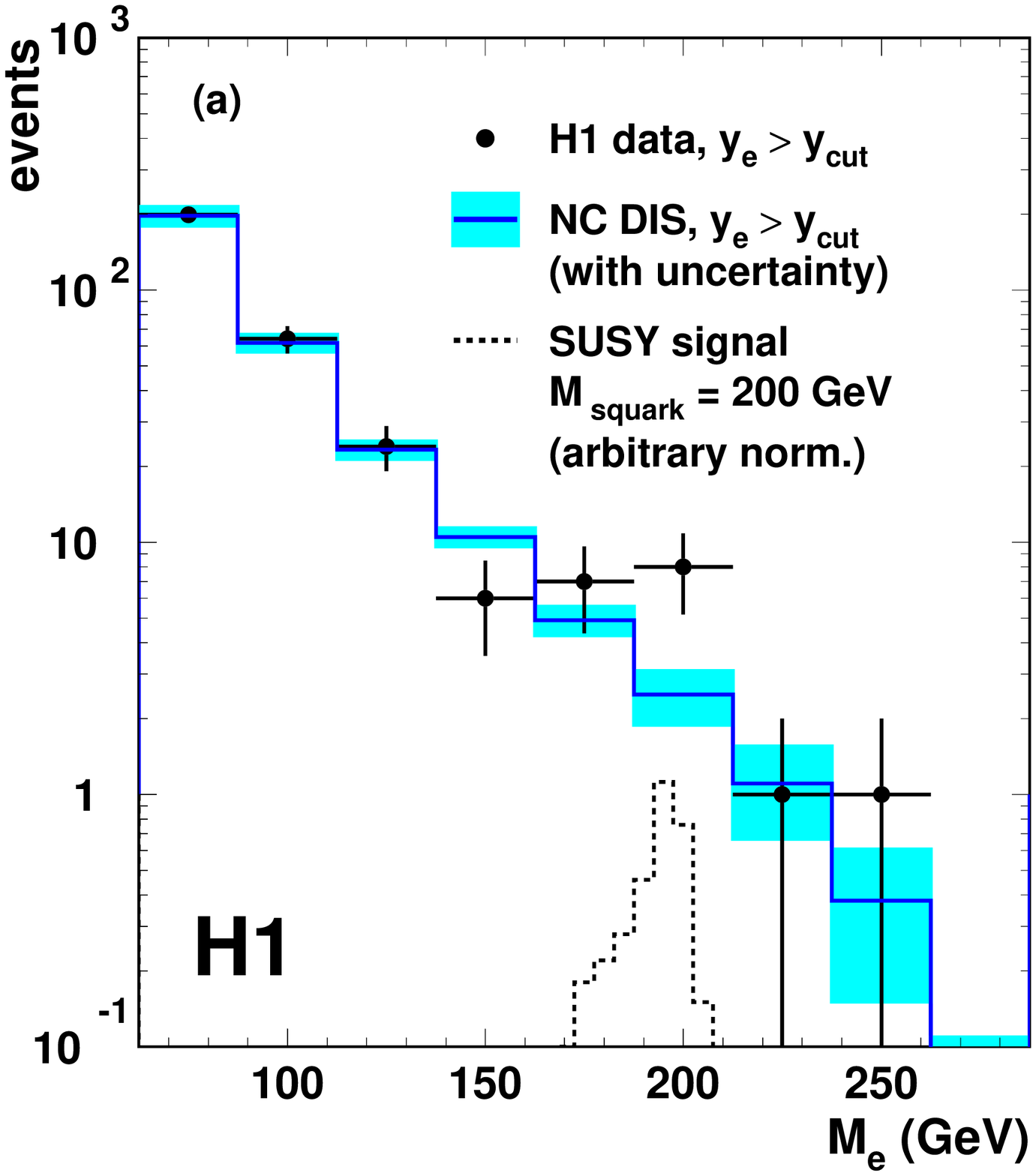}}
   &
     \hspace*{-0.8cm}\mbox{\epsfxsize=0.55\textwidth
        \epsffile{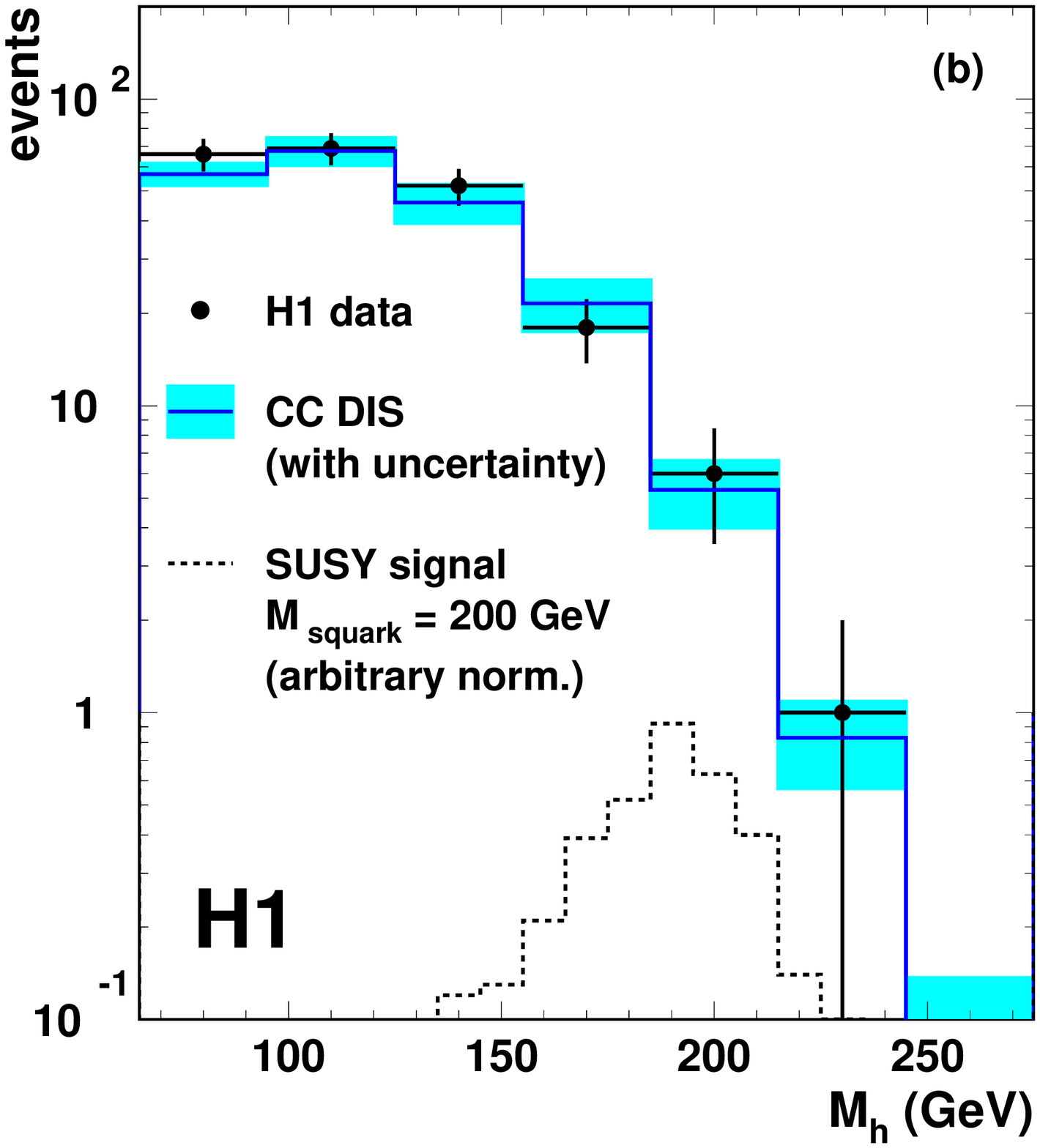}}
  \end{tabular}
  \end{center}
 \caption[]{ \label{fig:dndmS1S2}
 {\small 
         Mass spectra for (a) NC DIS-like and (b) CC DIS-like final states
         for data (symbols) and DIS expectation (solid histograms).
         In (a) the comparison is shown after a $M_e$ dependent
         cut on $y_e$ designed to maximise the significance of a
         squark signal~\cite{H1LQ99}.
         The grey boxes indicate the $\pm 1 \sigma$ band 
         of systematic errors of the DIS
         expectations.
         The dashed histograms show the mass distributions for
         simulated events coming from a $200 \GeV$ squark decaying into
         the channels (a) {\boldmath{$LQe$}} and
         (b) {\boldmath{$LQ \nu$}}, with an arbitrary
         normalisation. }}
\end{figure}

{\bf{Channel {\boldmath{$LQe$}}: }}
Squarks decaying into the channel {\boldmath $LQe$} have the same
signature as scalar leptoquarks and
are characterised by high $Q^2$ NC DIS-like
topologies. Such a process should manifest itself
as a resonance in the measured $M_e$ distribution,
with a resolution of $3$ to $6 \GeV$ depending on the squark mass.
The selection criteria are those described in~\cite{H1LQ99}.
The observed and expected mass spectra are shown
in Fig.~\ref{fig:dndmS1S2}a to be in good agreement,
with nevertheless a slight excess around $200 \GeV$
already reported in~\cite{H1LQ99,H1HIGHQ2}.
The sources of systematic errors are described
in section~\ref{sec:syst}.
The (arbitrarily normalised) 
mass distribution expected from signal events 
coming from a $200 \GeV$ squark decaying into the channel {\boldmath $LQe$}
is also shown.
The peak value is slightly
below the nominal squark mass due to final state
QCD radiation~\cite{H1LQ99}. 
Similar searches have been performed by the ZEUS experiment~\cite{ZEUSLQ99}.
%
%

 
\noindent
{\bf{Channel {\boldmath{$LQ\nu$}}: }} 
Squarks undergoing a {\boldmath $LQ\nu$} decay lead to CC DIS-like
events with high missing transverse momentum 
showing a clustering in the $M_h$ distribution with
a resolution of about $10 \%$ of the squark mass.
Only $\overline{\tilde{d}^k_R}$ squarks,
produced via a fusion between the incident $e^+$ and a $\bar{u}^j$ quark
can undergo such a decay.
The search for squarks decaying into the channel {\boldmath{$LQ\nu$}}
is described in~\cite{H1LQ99}.
The observed and expected mass
spectra are shown to be in good agreement in Fig.~\ref{fig:dndmS1S2}b.
The (arbitrarily normalised)
mass distribution expected from signal events
coming from a $200 \GeV$ squark decaying into the channel {\boldmath $LQ \nu$}
is also shown.
Similar searches have been performed by the ZEUS
experiment~\cite{ZEUSLQCC}.

 
\subsection{Analysis of Squark Gauge Decays 
            leading to {\boldmath{ $e$ + jets + $X$}} Topologies}
\label{sec:e+jets}

When the squark undergoes a gauge decay leading to a
positron, the final states can be classified
into several topologies, namely {\boldmath $eM\!J$}, 
{\boldmath $e^-\!M\!J$},
{\boldmath $eeM\!J$}, {\boldmath $e\mu M\!J$} and {\boldmath $\nu e  M\!J$}.
The ``$e$-preselection" requirements which are common to all these
$e$ + multijet channels are the following:
%
\begin{figure}[h]
 \hspace*{-0.6cm}\begin{tabular}{p{0.65\textwidth}p{0.35\textwidth}}
     \raisebox{-200pt}{
    \mbox{\epsfxsize=0.60\textwidth
       \epsffile{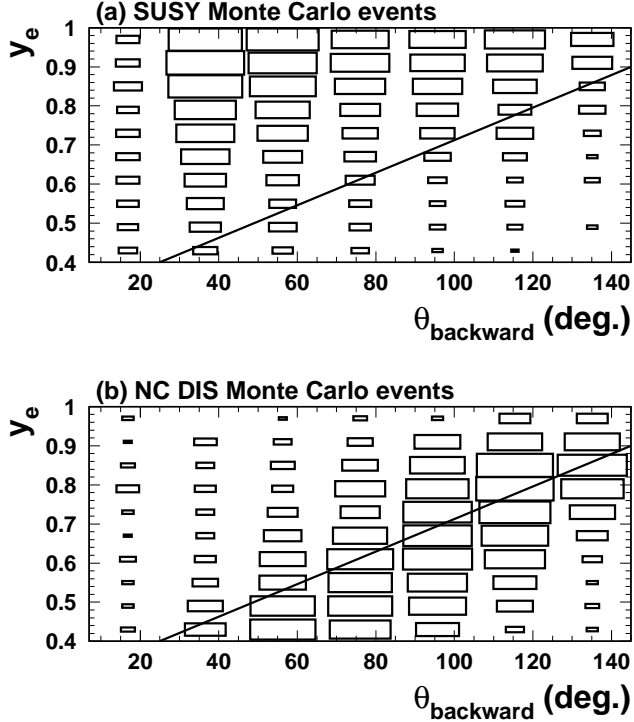}}}
 &
         \caption
         {\small \label{fig:thetacut}
         Correlation between $y_e$
         and the angle of the
         most backward jet, $\theta_{backward}$, for
         (a) SUSY Monte Carlo events where the squark undergoes
         a gauge decay leading to $e$ + multijets +$X$ final states
         and (b) NC DIS Monte Carlo
         events, when preselection cuts (1) to (3) are applied.
         Cut (4) only retains events above the diagonal  line.
         In (a) events were generated for the range of
         masses considered in this analysis. }
 \end{tabular}
\end{figure}
%
%
\begin{enumerate}
   \item at least one positron candidate in the angular range
         $5^{\circ} < \theta_e < 110^{\circ}$
         with $E_{T,e} > 5 \GeV$;
   \item at least two jets in the angular range
         $7^{\circ} < \theta < 145^{\circ}$;
         the highest $E_T$ jet must sa\-tis\-fy
         $\theta_{jet \, 1} > 10^{\circ}$ and $E_{T, \, jet \, 1} > 15 \GeV$;
         the second highest $E_T$ jet must have
         {\mbox{$E_{T, \, jet \, 2} > 10$ GeV}};
   \item  $y_e > 0.4$;
   \item of the two highest $E_T$ jets, 
         the one with the larger polar angle, $\theta_{backward}$,
         must satisfy:
         $$ y_e - 0.4 > (\theta_{backward}
           - 25^{\circ}) /60^\circ;$$
   \item the minimum of the polar angles of the highest $E_T$ positron and
      of the two highest $E_T$ jets must satisfy:
    $$\rm{Min}( \theta_e, \theta_{jet \, 1}, 
      \theta_{jet \, 2}) < 45^\circ \;\; . \;\;$$
\end{enumerate}

In gauge decays of a squark, a positron can emerge from the decay
of a $\chi$ or $\tilde{g}$ appearing in the decay chain.
It takes away a (possibly small) fraction of the momentum of this fermion,
which motivates the cut (3).
Moreover, 
it is strongly boosted in the direction of the incident proton,
such that the {\mbox{$\theta_e < 110^{\circ}$}}
requirement discriminates 
the signal from the NC DIS background.
Cut (4) exploits the fact that for high $y_e$ NC DIS events 
satisfying the above set of cuts, 
one hard jet is usually scattered in the
backward region of the calorimeter.
In contrast,
jets coming from a squark gauge decay will be boosted in the
forward direction. 
The $\theta_{backward}$ distribution for SUSY events depends
on the masses of the sparticles involved and cut
(4) was designed to always retain a large fraction of the
signal events.
The effect of cut (4)
is illustrated in Fig.~\ref{fig:thetacut}.
Cut (5) requires that one of the squark decay products should
be emitted in the forward direction and allows an additional
reduction of the SM background by
$\sim 40 \%$, with a negligible
efficiency loss on the signal.

%
\begin{figure}[htb]
  \begin{center}
     \mbox{\epsfxsize=0.8\textwidth
       \epsffile{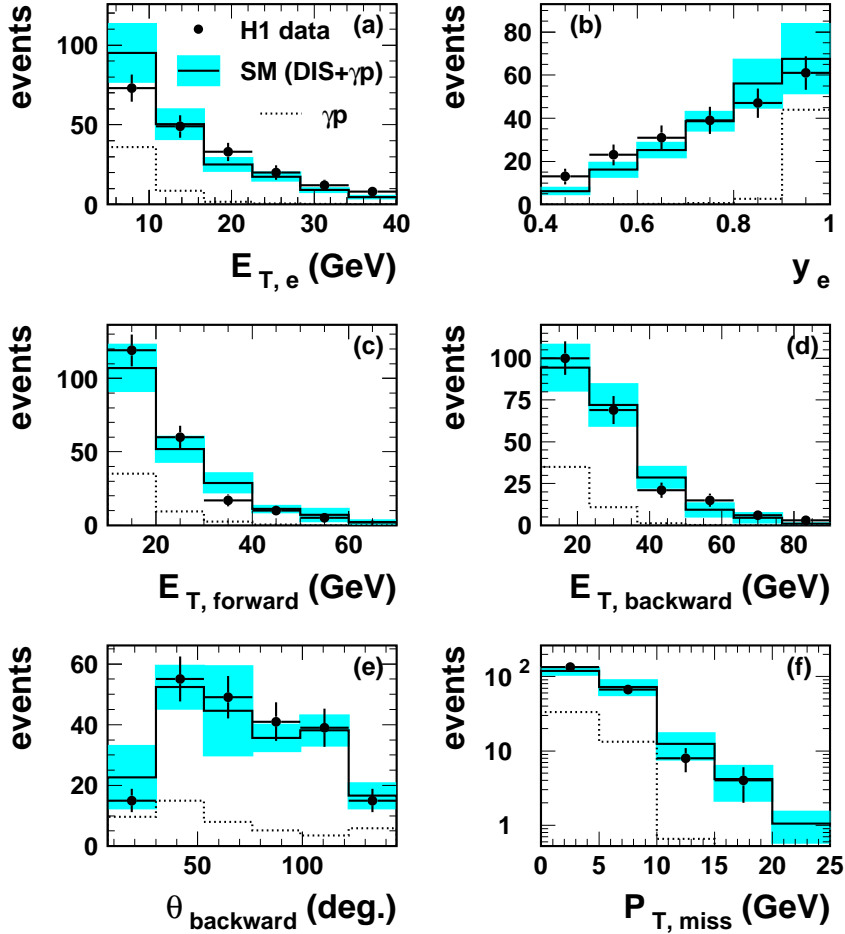}}
  \end{center}
 \caption[]{ \label{fig:nc_control}
 {\small For the $e$ + multijets $+X$ preselection, distributions of
      (a) the transverse energy $E_{T,e}$ of the highest $E_T$ positron;
      (b) $y_e$; (c) the transverse energy of the most forward jet;
      (d) the transverse energy of the most backward jet;
      (e) the polar angle of the most backward jet;
      (f) the missing transverse momentum. Superimposed on the data points
      (symbols) are histograms of the SM expectation (DIS and $\gamma p$).
      The grey band
      indicates the uncertainty on the SM prediction. The contribution
      from $\gamma p$ processes alone is shown as 
      dotted histograms.}}
\end{figure}

\begin{figure}[bht]
  \begin{center}
     \mbox{\epsfxsize=0.7\textwidth
        \epsffile{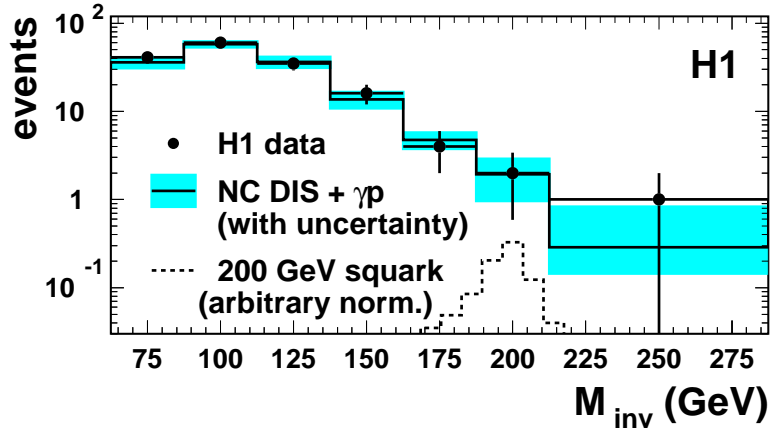}}
  \end{center}
 \caption[]{ \label{fig:dndmS3}
 {\small Mass spectrum for $e$ + multijet topology
      {\boldmath{$eM\!J$}} for the data
      (symbols) and the SM expectation (solid histogram).
      The grey band indicates the uncertainty on the SM prediction.
      The dashed histogram shows the expected mass distribution for
      events coming from a $200 \GeV$ squark decaying into
      the channel {\boldmath{$eM\!J$}}, with an arbitrary
      normalisation.}}
\end{figure}
%

Applying  the above selection criteria, 214 events are accepted,
which is in good agreement with the SM prediction of 
$210 \pm 34$, 
including 47 events from photoproduction
where a jet has been misidentified as an electron.
Fig.~\ref{fig:nc_control} shows the observed distributions
of the transverse energy of the highest $E_T$ positron, $y_e$,
the transverse energy of the most forward and most backward jet,
the polar angle of the most backward jet and the missing transverse
momentum. All distributions are seen to agree well with the
SM expectation within the systematic errors.

For channels leading to
$e + {\mbox{ jets }} + X$ final states additional cuts,
listed in table~\ref{tab:Sicuts}, 
are applied on top of the preselection requirements (1) to (5).
In each case, good agreement is observed between the data
and the SM expectation largely dominated by the NC DIS
contribution.
Additional information for the different channels
is given below. \\
 
%
\noindent{\bf Channel {\boldmath $eM\!J$: } }
%
A mass $M_{inv}$ is calculated as:
$$M_{inv}=\sqrt{4x_e E^0_e \left( \sum_{i} E_i - E^0_e \right) } \;\; , \; \;$$
where the sum runs over all energy deposits in the 
calorimeters for $\theta > 10^{\circ}$, thereby
excluding the proton remnant.
For squarks decaying into the {\boldmath{$eM\!J$}} channel, $M_{inv}$ provides
an estimate of the $\tilde{q}$ mass.
This reconstruction method yields a typical resolution
of $7$ to $10 \GeV$ depending on the squark mass.
The $M_{inv}$ spectrum of the selected events is 
shown in Fig.~\ref{fig:dndmS3} to be well described by the SM prediction.
Also shown is the (arbitrarily normalised)
mass distribution expected from signal events
coming from a $200 \GeV$ squark decaying into the channel 
{\boldmath  $eM\!J$}.
No charge determination of the lepton
is performed here.

\noindent{\bf Channel {\boldmath $e^-\!M\!J$: }} 
%
We consider the track  
in the $e$ isolation cone which has the highest
momentum projected on the axis defined by the event vertex and
the centre of gravity of the calorimetric energy deposits
associated with the electron.
This track is required to have a reliably measured negative charge.
The efficiency 
of the  track quality requirements is $\simeq 80 \%$,
derived from data (candidates for channel {\boldmath{$eM\!J$}} ) 
and well reproduced by the NC DIS simulation.

\noindent{\bf Channel {\boldmath $e \ell M\!J$: } } 
%
Di-lepton final states are searched for
provided that the lepton accompanying the $e$
belongs to the first or second generation. 
%
\begin{figure}[h]
 \begin{center}
 \hspace*{-0.6cm}\begin{tabular}{p{0.57\textwidth}p{0.43\textwidth}}
     \raisebox{-170pt}{
    \mbox{\epsfxsize=0.5\textwidth
       \epsffile{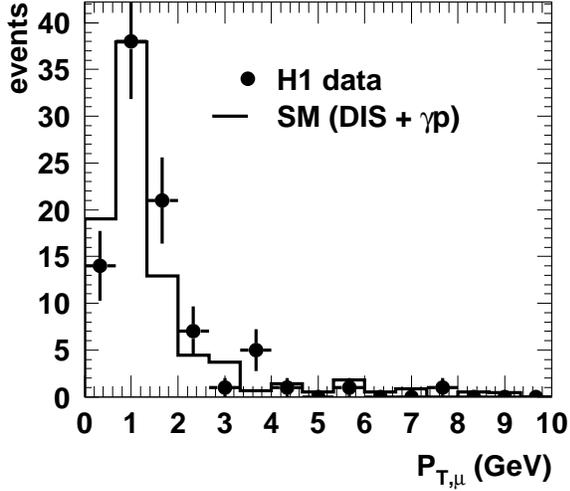}}}
 &
        \caption[]{ \label{fig:ptmuon}
       {\small Distribution of the transverse momentum $P_{T,\mu}$
        of the muon for events satisfying the 
        preselection criteria and where a muon has
        been identified
        in the angular range $10^{\circ} < \theta < 110^{\circ}$,
        for data (points) and SM (histogram).
       }} 
 \end{tabular}
 \end{center}
\end{figure}
%
For the channel {\boldmath{$e \mu M\!J$}},
the properties of the muons observed in the preselected events
over the full range in transverse momentum
were found to be well described by the simulation,
as exemplified in Fig.~\ref{fig:ptmuon}.

\begin{figure}[hbt]
  \begin{center}
     \mbox{\epsfxsize=0.7\textwidth
       \epsffile{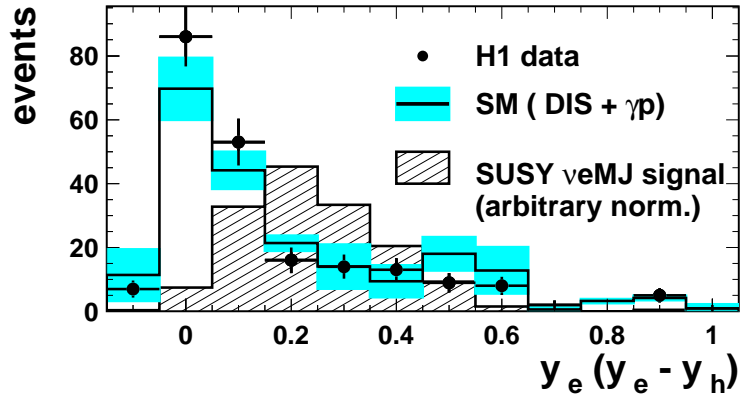}}
  \end{center}
 \caption[]{ \label{fig:s8ecut}
 {\small Distribution of the variable $y_e \cdot (y_e - y_h)$
       for the 214 events satisfying the preselection criteria
       (symbols) and for the expectation from NC DIS and $\gamma p$ 
       processes (open histogram). 
       The grey error band on the
       open histogram indicates the uncertainty on the SM prediction.
       The arbitrarily normalised hatched histogram shows how this
       variable is distributed for events coming from a squark
       decaying into the channel {\boldmath{ $\nu eM\!J$}},
       generated for the range of masses considered in this
       analysis. 
      }}
\end{figure}

\noindent{\bf Channel {\boldmath $\nu eM\!J$: }}
%
The common preselection criteria are complemented by a 
$P_{T,miss}$ requirement and by the cut 
$y_e \cdot (y_e - y_h) > 0.05$.
A cut on the product of $y_e$ with the difference
$(y_e - y_h)$
exploits the fact that, for events coming from
a squark decaying into the channel 
{\boldmath{ $\nu eM\!J$}}, the escaping neutrino
carries  a non-negligible part of 
$\sum \left( E-P_z \right) $ and hence
the variable $y_h$ is substantially smaller than $y_e$,
while $y_e \sim y_h$ is expected for NC DIS events.
Fig.~\ref{fig:s8ecut} shows the distribution of
$y_e \cdot (y_e - y_h)$ for the 214 events accepted by the
preselection and for the SM expectation.
%
%

\subsection{Analysis of Squark Gauge Decays 
            leading to {\boldmath{ $\nu$ + jets + $X$}} Topologies}


Two channels are considered to cover cases where squarks undergo 
a gauge decay leading to a neutrino (and no positron)
in the final state.
These final
states, {\boldmath{$\nu M\!J$}} and {\boldmath{$\nu\mu M\!J$}},
are selected by the following
``$\nu$-preselection" requirements:
\begin{enumerate}
\item
       a missing transverse momentum $P_{T,miss} > 25$~GeV;
\item at least two jets in the angular range
       $7^{\circ} < \theta < 145^{\circ}$ and with $E_T > 10 \GeV$, with
       the highest $E_T$ jet  satisfying
       $\theta_{jet \, 1} > 10^{\circ}$ and $E_{T, \, jet \, 1} > 15 \GeV$. 
\end{enumerate}
%
%
\begin{figure}[htb]
  \begin{center}
     \mbox{\epsfxsize=0.79\textwidth
       \epsffile{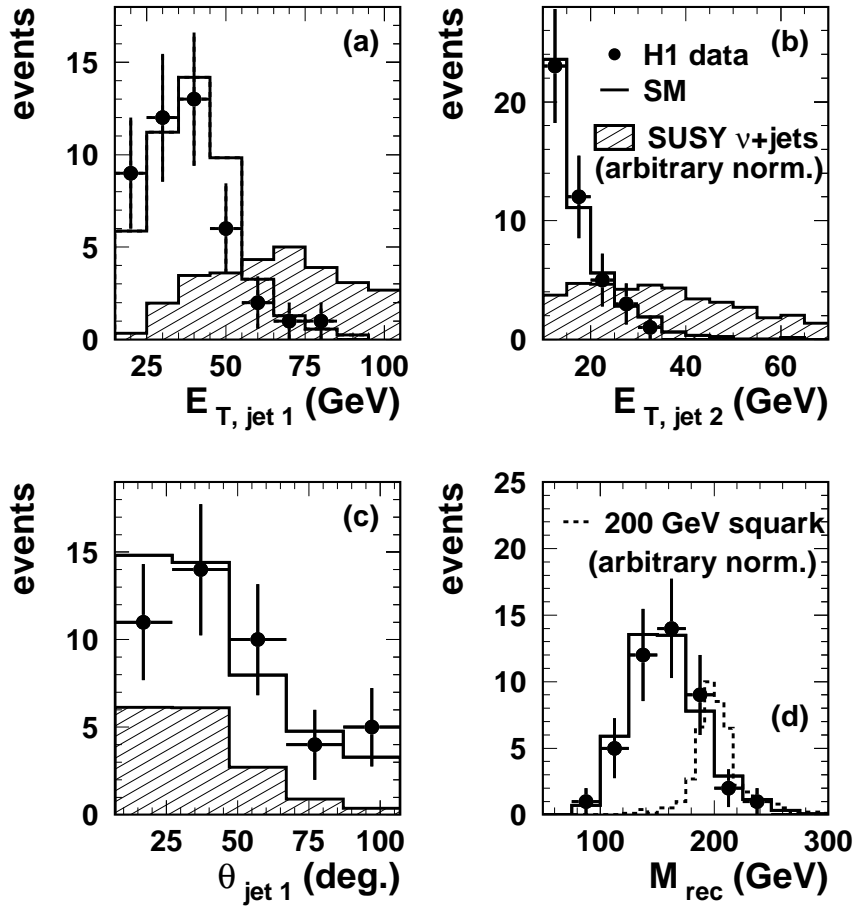}}
  \end{center}
 \caption[]{ \label{fig:cc_control}
 {\small For the $\nu$ + multijets $+X$ preselection, 
      observed (symbols) and expected (open histograms) distributions of
      (a), (b) the transverse energies of the two highest $E_T$ jets 
      and (c) the polar angle of the highest $E_T$ jet.
      The hatched histograms show distributions  
      for events from a 
      squark gauge decay into {\mbox{$\nu$ + jets}},
      generated for the range of masses considered in this analysis.
      (d): The mass $M_{rec}$, corresponding to the energy
          in the centre of mass of the hard
          subprocess assuming that only one neutrino escapes detection;
          the dashed histogram shows the $M_{rec}$ distribution for
          signal events coming from the decay of a $200 \GeV$ squark into the
          channel {\boldmath{$\nu M\!J$}}.
          All histograms showing the SUSY  expectation are
          arbitrarily normalised. }}
\end{figure}
%
%
We observe 44 events satisfying these preselection criteria, in good
agreement with the SM prediction of $46.5 \pm 6.9$, mainly coming
from CC DIS processes.

%
\begin{table}[hp]
 \renewcommand{\doublerulesep}{0.8pt}
 \renewcommand{\arraystretch}{1.0}
 \begin{center}
  \begin{tabular}{||c|c|c|c|c||}
  \hline \hline
  {\bf Channel}  &  \multicolumn{1}{c|}{\bf Selection Cuts}
           & \multicolumn{1}{c|}{\bf  Efficiency} 
           & \multicolumn{1}{c|}{\boldmath $N_{obs}$ }
           & \multicolumn{1}{c||}{\boldmath $N_{exp}$}         \\ 
  & & & & \\
\hline
%
 \multicolumn{5}{||c||}{ {\bf \boldmath
DIS-like channels:} $Q^2 > 2500$ GeV$^2$, $y < 0.9$} \\ \hline \hline
 & & & & \\
  {\boldmath{$LQe$}} &  \begin{tabular}{c}
        $E_{T,e} > 15$ GeV \\
        $P_{T,miss}/\sqrt{E_{T,e}} \leq 4 \sqrt{\GeV}$ \\
        $ 40 \leq \sum \left(E - P_z\right) \leq 70 \GeV$ \\
        optimised lower $y$-cut
        \end{tabular} 
     &  \begin{tabular}{c}
        40--70 $\%$
        \end{tabular} 
     &  310
     &  301 $\pm$ 23 \\ \hline
 & & & & \\
  {\boldmath{$LQ\nu$}} &  \begin{tabular}{c}
        $P_{T,miss} > 30 {\GeV}$ \\
        no electron $E_{T,e}>5 {\GeV}$ 
        \end{tabular} 
     &  \begin{tabular}{c}
        30--80 $\%$
        \end{tabular} 
     &  213
     &  199 $\pm$ 12 \\ 
 & & & & \\
\hline \hline
 \multicolumn{5}{||c||}{\bf \boldmath 
channels with: $e$ + multijets + X} \\ 
 \multicolumn{5}{||c||} { $e$-preselection:  $E_{T,e}>5\GeV$;
$\geq$ 2 jets: $E_{T, \, jet \, 1,2}>15,10 \GeV$; high $y_e$; angular cuts} \\

\hline \hline
 & & & & \\
  {\boldmath{$eM\!J$}} &  \begin{tabular}{c}
        $P_{T,miss} < 20 {\GeV}$ \\
        $ 40 \leq \sum \left(E - P_z\right) \leq 70 \GeV$ \\
        \end{tabular} 
     &  \begin{tabular}{c}
        35--50 $\%$
        \end{tabular} 
     &  159
     &  151 $\pm$ 17 \\ 
 & & & & \\
\hline 
 & & & & \\
  {\boldmath{$e^-\!M\!J$}} &  \begin{tabular}{c}
        $eM\!J$ criteria \\
        + ``wrong" charge of $e$
        \end{tabular} 
     &  \begin{tabular}{c}
        $\approx 30\%$
        \end{tabular} 
     &  0
     &  1.3 $\pm$ 0.5 \\ 
 & & & & \\
\hline 
 & & & & \\
  {\boldmath{$eeM\!J$}} &  \begin{tabular}{c}
         second $e$ with: \\
         $E_{T,e2} > 5$ GeV  \\
         $5^{\circ} < \theta_{e2} < 110^{\circ}$
        \end{tabular} 
     &  \begin{tabular}{c}
        $\approx 30\%$
        \end{tabular} 
     &  0
     &  0.7 $\pm$ 0.4 \\ 
 & & & & \\
\hline
 & & & & \\
   {\boldmath{$e \mu M\!J$}} &  \begin{tabular}{c}
         $P_{T,\mu} > 5$  GeV \\
         $10^{\circ} < \theta_{\mu} < 110^{\circ}$
        \end{tabular} 
     &  \begin{tabular}{c}
        35--50$\%$
        \end{tabular} 
     &  2
     &  4.2 $\pm$ 1.2 \\ 
 & & & & \\
\hline
 & & & & \\
  {\boldmath{$\nu eM\!J$}} &  \begin{tabular}{c}
        $P_{T,miss} > 15 \GeV$  \\
        $ y_e (y_e - y_h) > 0.05$ \\
        \end{tabular} 
     &  \begin{tabular}{c}
        $\approx 30\%$
        \end{tabular} 
     &  1
     &  3.2 $\pm$ 1.2 \\ 
 & & & & \\
\hline   \hline

 \multicolumn{5}{||c||}{\bf channels with: {\boldmath{$\nu$}} + multijets + X} \\ 
  \multicolumn{5}{||c||}{$\nu$-preselection: $P_{T,miss} > 25\GeV$; $\geq$ 
2 jets: $E_{T, \, jet \, 1,2}>15,10\GeV$}  \\
\hline \hline
 & & & & \\
  {\boldmath{$\nu M\!J$}} &  \begin{tabular}{c}
        $E_{T,jet2} > 15 {\GeV}$ \\
        $\sum \left( E-P_z \right)_{h} < 55$ GeV \\ 
        \end{tabular} 
     &  \begin{tabular}{c}
        20--60 $\%$
        \end{tabular} 
     &  21
     &  23 $\pm$ 4 \\ 
 & & & & \\
\hline 
 & & & & \\ 
  {\boldmath{$\nu \mu M\!J$}} &  \begin{tabular}{c}
        $P_{T,\mu} > 5 {\GeV}$ \\
         $10^{\circ} < \theta_{\mu} < 110^{\circ}$
        \end{tabular} 
     &  \begin{tabular}{c}
        $\approx 40 \%$
        \end{tabular} 
     &  0
     &  0.5 $\pm$ 0.2 \\ 
 & & & & \\
\hline 
%
   \hline \hline
  \end{tabular}
  \caption[]
          {\small \label{tab:Sicuts}
           Selection criteria, typical efficiencies,
           total number of observed events $N_{obs}$ and the corresponding
           SM expectation $N_{exp}$ with its uncertainty, for each
           squark decay channel analysed. 
           The ``e-preselection" criteria are detailed in 
           section~\ref{sec:e+jets}.
          }
 \end{center}
\end{table}
Figs.~\ref{fig:cc_control}a-c show
the distributions of
the transverse energies of the two highest $E_T$ jets and of the polar angle
of the highest $E_T$ jet. The data are well described by
the SM expectation. 
Distributions for squarks decaying into
{\mbox{$\nu$ + jets}} are also shown.
Assuming that the missing energy is carried by one neutrino only,
its kinematics is reconstructed exploiting
energy-momentum conservation.
The four-vector of this $\nu$ is then added to that of
the hadronic final state to reconstruct the invariant mass $M_{rec}$ 
of the incoming electron and quark. 
For squarks decaying
into the channel {\boldmath $\nu M\!J$}, $M_{rec}$ provides an estimate of
the squark mass.
The observed and expected distributions for $M_{rec}$ are in good agreement
as shown in Fig.~\ref{fig:cc_control}d.
The dashed, arbitrarily normalised, 
histogram in Fig.~\ref{fig:cc_control}d shows
the resulting mass spectrum for a $200 \GeV$
squark decay into the {\boldmath $\nu M\!J$} channel.
The observed  resolution of $\sim 15 \GeV$
is typical for the range of squark masses probed in this analysis.

Final cuts and results for the channels {\boldmath $\nu M\!J$}
and {\boldmath $\nu \mu M\!J$} are given in table~\ref{tab:Sicuts}.
The number of observed candidates is in good agreement with the
SM expectation, which is dominated by the contribution of CC DIS processes.

 
\subsection{Systematic Errors}
\label{sec:syst}

In each channel,
the error on
the expectation from Standard Model processes has been calculated
by taking into account the systematic errors described below. 
The experimental error sources considered are:
\begin{itemize}
 \item an uncertainty of $\pm 1.5 \%$ on the integrated luminosity; 
 \item an uncertainty on the absolute calibration of the calorimeters
       for electromagnetic energies, ranging between $\pm 0.7\%$ in the
       central part and $\pm 3\%$ in the forward region of the LAr
       calorimeter; this constitutes the main error source for
       the estimation of NC DIS background to the channel {\boldmath{$LQe$}};
 \item an uncertainty of $\pm 4 \%$ on the absolute hadronic energy
       scale. For inclusive NC DIS final states,
       the over-constrained kinematics allows a reduction of this
       uncertainty to $\pm 2 \%$~\cite{H1F2PAPER}, which applies to
       channels {\boldmath{$LQe$}} and {\boldmath{$LQ\nu$}}.
       This is the main error source for all  channels
       except {\boldmath{$LQe$}}.
\end{itemize}
The following theoretical
uncertainties on SM cross-sections are considered.
\begin{itemize}
 \item For NC DIS-like final states, an 
       uncertainty of $\pm 5 \%$ is attributed to the
       proton structure~\cite{H1HIGHQ2}, which is
       partly due to the experimental errors on the input data 
       entering the QCD fits, and partly linked to the assumptions
       for the shapes of the parton distributions at the scale 
       where the perturbative QCD evolution is started.
       For CC DIS-like topologies,
       which are mainly induced by $d$ quarks
       whose density in the proton is less constrained, this
       uncertainty increases linearly with $Q^2$ up to $\simeq 20 \%$
       at the highest $Q^2$ considered here~\cite{H1LQ99}.
       In adition, the error on the strong coupling constant $\alpha_S$
       leads to an uncertainty of $\pm 4\%$ on the proton structure.
       This was inferred~\cite{H1HIGHQ2} by comparing the CTEQ4 (A1) to (A5)
       parametrisations~\cite{CTEQA1A5} with $\alpha_S (M_Z)$ ranging
       between 0.110 to 0.122.
 \item 
       Higher order QED corrections imply a $\pm 2 \%$ uncertainty
       in the $y$ range considered here~\cite{H1HIGHQ2},
       estimated using the HECTOR~\cite{HECTOR} program.
 \item An uncertainty of $\pm 10 \%$ on the predicted cross-section
       for multijet final states is estimated by comparing leading
       order (LO) Monte Carlo simulations
       where higher order QCD radiation is modelled by
       either the dipole model or DGLAP parton showers,
       and next-to-leading order 
       (NLO) calculations.
\end{itemize}
For each error source, the analysis has been repeated  shifting
the central value  by $\pm 1$ standard deviation to estimate their
individual contribution.
The overall systematic error on SM expectations is then determined as
the quadratic sum of these individual errors and the statistical uncertainty
on the Monte Carlo simulation. \\

\section{Constraints on SUSY Models }
\label{sec:results}

No significant deviation from  SM expectations 
has been found in the analysis of various $\Rp$ and gauge 
decay channels.
These channels are
combined to set constraints on \Rp\ SUSY models.

As mentioned in section~\ref{sec:pheno},
when HERA operates with incident positrons,
the best probed \Rp\ couplings are $\lambda'_{1j1}$, on which we mainly
concentrate here. 
Such a coupling allows for the production of $\tilde{u}^j_L$
or $\overline{\tilde{d}^k_R}$ squarks, the latter with a much reduced rate
due to the smaller parton density.
Thus only $\tilde{u}^j_L$ production is considered.
Since this squark does not decay into $\nu + q$
the channel {\boldmath{$LQ\nu$}} is not taken into account in
the derivation of limits.
The $\tilde{s}_R$ production is also (conservatively) neglected when
setting limits on $\lambda'_{1j2}$.

In this section, mass dependent upper limits
on the Yukawa couplings are first derived
in a ``phenomenological'' version of the MSSM
where the masses of the sfermions are not related to the SUSY 
soft-breaking mass terms of the gauginos.
Scans are then performed in the framework of this ``phenomenological''
MSSM as well as in the constrained MSSM, and 
bounds on the Yukawa couplings are set in these models. 
Results are finally interpreted in the framework of the 
minimal Supergravity model.

\subsection{Derivation of Limits}
\label{sec:limideri}

Mass dependent upper limits on the production cross-section
$\sigma(e^+ p \rightarrow \tilde{u}^j_L)$ 
are obtained assuming Poisson distributions
for the SM background expectations as well as for the signal.
A standard Bayesian prescription with a flat prior
probability density for the signal cross-section is used.
Both systematic and statistical errors have been folded in
channel by channel as described in~\cite{H1RPVEMINUS}.
Each channel contributes in the derivation of the limits
via its branching ratio,
the number of observed and expected events satisfying
the relevant selection cuts and the corresponding selection efficiency.
For the channels {\boldmath{$LQe$}}, {\boldmath{$eM\!J$}} and 
{\boldmath{$\nu M\!J$}}, the numbers of
observed and expected events are integrated within a mass bin 
which slides over the accessible mass range.
The width of the mass bin is adapted to the expected mass resolution
in each channel, such that this bin contains approximately
$68 \%$ of the signal at a given squark mass.
For the channels {\boldmath{$e^-\!M\!J$}}, {\boldmath $e \ell M\!J$} 
and {\boldmath{$\nu \ell M\!J$}},
where both the SM expectation and the observation
are small, no mass restriction is imposed.

%
Events which fulfil the selection requirements
of more than one channel are only counted as squark candidates in the 
channel with the highest sensitivity. This prescription is
illustrated in table~\ref{tab:exclucuts}.
Note that the potential
overlap between channels {\boldmath{$LQe$}} and {\boldmath{$eM\!J$}}
is already considerably reduced by the mass bin requirements,
since the corresponding reconstructed squark masses,
$M_e$ and $M_{inv}$ respectively, differ
by typically more than $20 \GeV$.
The relative efficiency loss on the SUSY signal induced by
this slight tightening of the cuts depends on the
masses of the sparticles involved, and is found to 
vary between $\sim 0.5 \%$ and $\sim 3 \%$.

\begin{table}[hbt]
 \begin{center}
  \begin{tabular}{c|c}
 candidates fulfilling the selection criteria for channels:  &
  are not considered in channel: \\
  \hline 
& \\
{\boldmath{$eM\!J$}} 
\hspace{0.6cm} {\boldmath{$e^-\!M\!J$}} \hspace{0.6cm}
{\boldmath{$e e M\!J$}}\hspace{0.6cm} 
{\boldmath{$e \mu M\!J$}} \hspace{0.6cm} {\boldmath{$\nu eM\!J$}}
& {\boldmath{$LQe$}} \\

{\boldmath{$e^-\!M\!J$}}\hspace{0.6cm}
{\boldmath{$e e M\!J$}}\hspace{0.6cm} 
{\boldmath{$e \mu M\!J$}} \hspace{0.6cm} {\boldmath{$\nu eM\!J$}}
& {\boldmath{$eM\!J$}} \\

{\boldmath{$eeM\!J$}}\hspace{0.6cm}
{\boldmath{$e\mu M\!J$}} & {\boldmath{$\nu eM\!J$}} \\
{\boldmath{$\nu \mu M\!J$}}  & {\boldmath{$\nu  M\!J$}} \\
\end{tabular}
\caption[]{\small \label{tab:exclucuts}
{\small   
     Prescription adopted to ensure no overlap
     between the considered channels. }}
   \end{center}
\end{table}

The masses of the neutralinos, charginos and gluinos, as
well as the couplings between any two SUSY particles and a standard
fermion/boson, are  
determined by the usual MSSM parameters:
the ``mass'' term $\mu$ which mixes the Higgs superfields, the 
SUSY soft-breaking
mass parameters $M_1$, $M_2$ and $M_3$ for $U(1)$, $SU(2)$ 
and $SU(3)$ gauginos,
and the ratio $\tan \beta$
of the vacuum expectation values of the two neutral scalar Higgs fields.
These parameters are defined at the electroweak (EW) scale. 
We assume that the gaugino mass terms unify at a
Grand Unification (GUT) scale to a
common value $m_{1/2}$ leading to usual relations~\cite{MSSM}
between $M_1$, $M_2$ and
$M_3$, and
approximate the gluino mass by the value of $M_3$ at the EW scale.
The masses and decay widths of all involved sparticles
have been obtained using the SUSYGEN package.

For a fixed set of MSSM parameters and 
sfermion masses,
the branching ratios for the different channels
only depend on the Yukawa coupling.
This also holds for the upper limit on the signal cross-section
$\sigma_{lim}$ derived from the combination of the analysed channels.
At a given squark mass, Yukawa couplings for which
$\sigma_{lim}$ is smaller than the signal cross-section
are excluded, where the signal cross-section is obtained by
multiplying the leading-order production cross-section
by $K$-factors~\cite{SPIRANLO} accounting for NLO QCD effects.
These can enhance the signal rate by ${\cal{O}}(10 \%)$.

Decay chains involving more than two SUSY fermions ($\chi$ or $\tilde{g}$) 
can contribute in principle to the gauge channels analysed.
In these cases parameterising the signal efficiencies 
is not straightforward.
Hence, only cascades involving two SUSY fermions 
are taken into account in 
the calculation of the visible branching ratios for
gauge decay channels.
This determination of the branching ratios
is conservative. It has been checked that
the visible branching is
generally close to $100 \%$.
Decays of $\chi$'s into a Higgs boson
are included in the calculation of the visible branching ratios
when the Higgs decays into hadrons. 
The contribution of these decays is however very small.
Hence the limits do not depend on the mass $m_A$ of the pseudoscalar Higgs,
set here to $300 \GeV$ in the models where $m_A$ is not related to 
the other parameters.

The case of a non-vanishing coupling $\lambda'_{131}$ allowing
for the resonant production of a stop squark (SUSY partner
of the top quark) has to be treated
separately.
Firstly, the large top mass can not be neglected
in the calculation of the branching ratios for the decays
$\tilde{t}  \rightarrow t \chi^0$ or $\tilde{t}  \rightarrow t \tilde{g}$,
which may eventually be kinematically forbidden.
Secondly, the top quark decays via $t \rightarrow b W$.
Most of the stop decays are in fact covered by our analysis,
but the efficiencies for the considered channels, which
are valid for decay patterns as shown in table~\ref{tab:sqtopo1},
can not be used in that case.
Conservatively, diagrams which lead to a top in the final
state are thus not taken into account
in the calculation
of the visible branching ratios.
For example, the \Rp\ decays $\chi^0_i \rightarrow e^- t \bar{d}$
will not be included in 
the branching ratios for the {\boldmath{$eM\!J$}} and
{\boldmath{$e^-\!M\!J$}} channels even when these are kinematically allowed.
As a result, only the neutrino decays of the $\chi^0_i$
($\chi^0_i \rightarrow \nu d \bar{d}$)
will contribute in the derivation of limits on $\lambda'_{131}$.

\subsection{Limits on {\boldmath $\lambda'_{1j1}$} and {\boldmath $\lambda'_{1j2}$} 
in the ``phenomenological'' MSSM}
\label{sec:unconstrained}

We consider here a version of the MSSM where
the parameters $\mu$, $M_2$ and $\tan \beta$ are only used to
determine the masses and couplings of the $\chi$'s, while
the sfermion masses are free parameters.
We neglect any possible mixing between sfermions 
and assume that all squarks are degenerate in mass.
This assumption only
enters in the calculation of the branching ratios of the $\chi$'s
and of the gluino, since we are mainly probing the $\tilde{u}^j_L$
squark. Sleptons are also assumed to be degenerate, and their mass
$M_{\tilde{l}}$ is set either to the common squark mass, or to a fixed value
($90 \GeV$) close to the lowest mass bound from sfermion
searches at LEP.
We first derive constraints on the couplings $\lambda'_{1j1}$,
where a squark could be produced via an $e^+ d$ fusion,
and consider in a
second step squark production via $e^+ s$ fusion through
a $\lambda'_{1j2}$ coupling.


\begin{figure}[b]
  \begin{center}
   \mbox{\epsfxsize=0.8\textwidth
      \epsffile{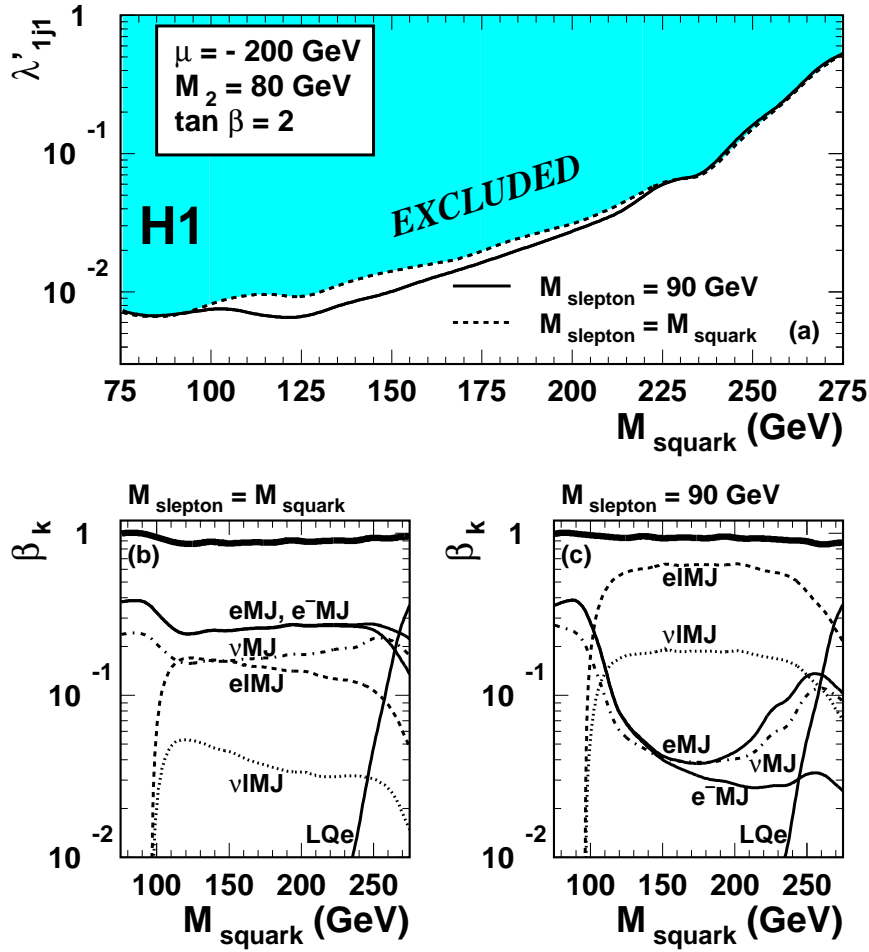}}
  \end{center}
 \caption[]{ \label{fig:lim_combine_photino}
 {\small  (a): Upper limits at $95 \%$ CL for the coupling
       $\lambda'_{1j1}$ ($j=1,2$) as a function of the squark mass, for a set
       of MSSM parameters leading to a $\chi^0_1$ 
       of $\sim 40 \GeV$ dominated
       by its photino component.
       Regions above the curves are excluded.
       The limits are given for two hypotheses on the slepton
       mass.
       (b): The branching ratios of all channels 
       versus the squark mass, when sleptons and squarks
       are assumed to be degenerate;
       (c): as (b) but assuming a slepton mass of $90 \GeV$. 
       The thick curves in (b) and (c) indicate the total
       branching covered by the analysis. }}
\end{figure}
%

\begin{figure}[b]
  \begin{center}
   \mbox{\epsfxsize=0.8\textwidth
      \epsffile{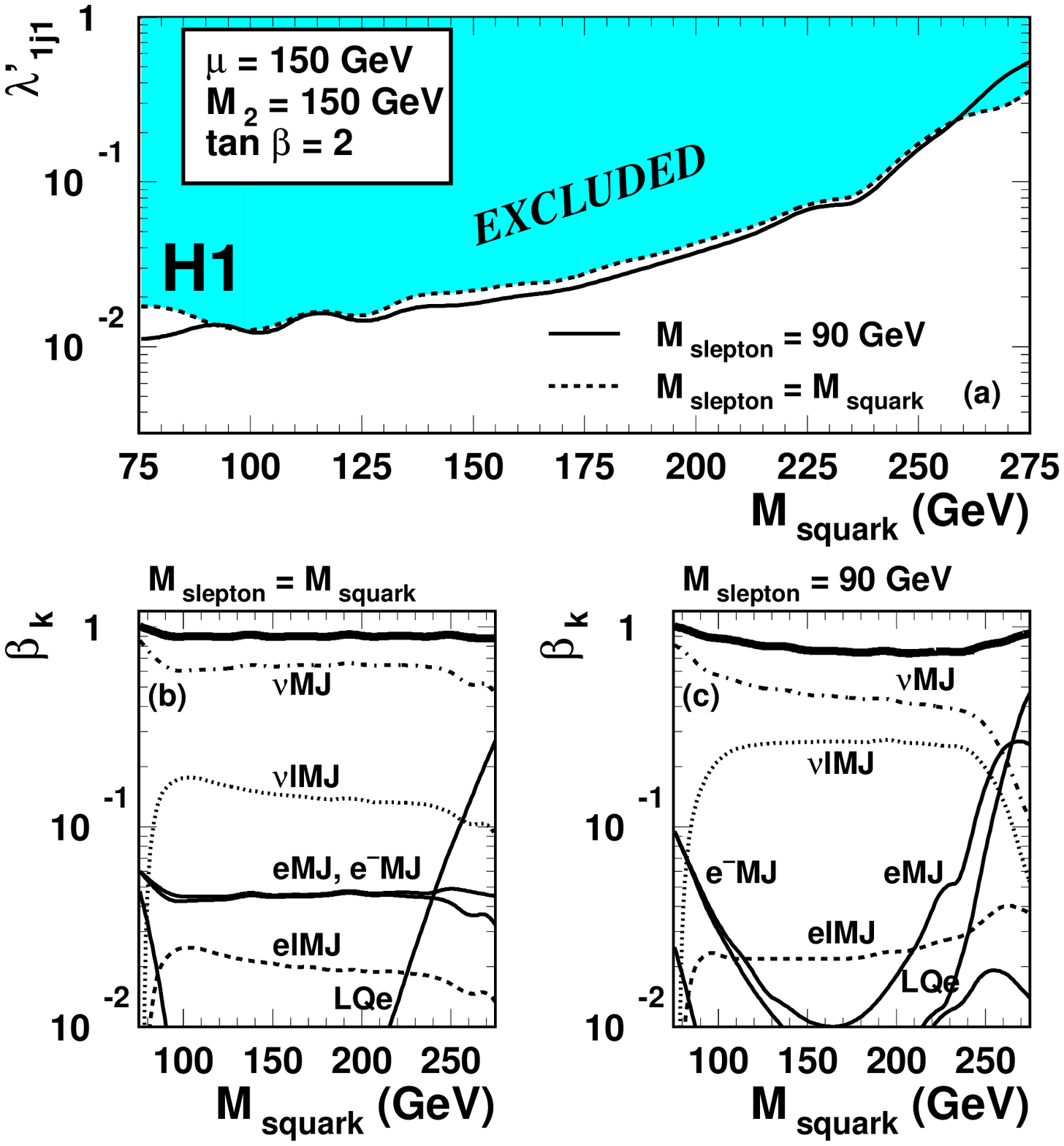}}
  \end{center}
 \caption[]{ \label{fig:lim_combine_zino}
 {\small  (a): Upper limits at $95 \%$ CL for the coupling
       $\lambda'_{1j1}$ $(j=1,2$) as a function of the squark mass, for a set
       of MSSM parameters leading to a $\chi^0_1$ 
       of $\sim 40 \GeV$ dominated
       by its zino component.
       Regions above the curves are excluded.
       The limits are given for two hypotheses on the slepton
       mass.
       (b): The branching ratios of all channels 
       versus the squark mass, when sleptons and squarks
       are assumed to be degenerate;
       (c): as (b) but assuming a slepton mass of $90 \GeV$. 
       The thick curves in (b) and (c) indicate the total
       branching covered by the analysis. }}
\end{figure}

Example upper limits obtained at  $95 \%$ confidence level (CL)
on $\lambda'_{1j1}$ ($j=1,2$) as a function of the $\tilde{u}^j_L$ mass are
shown in Fig.~\ref{fig:lim_combine_photino}a and 
Fig.~\ref{fig:lim_combine_zino}a, 
under the assumption $M_{\tilde{l}} = 90 \GeV$
or $M_{\tilde{l}} = M_{\tilde{q}}$.
The MSSM parameters are chosen such that the
lightest neutralino is dominated by its photino component in
Fig.~\ref{fig:lim_combine_photino} and by its zino component
in Fig.~\ref{fig:lim_combine_zino}.
The gluino mass is large
due to the large $M_2$ (and hence $M_3$) 
values and thus squark decays into $\tilde{g}$
are kinematically forbidden.
In the four scenarios considered, Yukawa couplings 
larger than $\sim 0.01$ ($\sim 0.5$)
are ruled out for squark masses of $75 \GeV$ $(275 \GeV)$.

The relative contributions of all channels 
are shown in Fig.~\ref{fig:lim_combine_photino}b,c
and Fig.~\ref{fig:lim_combine_zino}b,c
against the squark mass, at the current sensitivity 
limit on the Yukawa coupling.
In each case the total branching ratio covered by the
analysis is close to 100 $\%$.
At the largest squark masses, where a large Yukawa coupling is
necessary to allow squark production, the relative contribution of
the {\boldmath{$LQe$}} channel becomes important.
In the case illustrated in Fig.~\ref{fig:lim_combine_photino}
(Fig.~\ref{fig:lim_combine_zino}) the dominant channels
are those with an $e^{\pm}$ ($\nu$) in the final state, 
resulting in particular from the main decay mode of the $\chi^0_1$.
The relative contributions of the channels strongly depend
on the slepton mass. In the case  illustrated 
in Fig.~\ref{fig:lim_combine_photino}c, where
a small slepton mass $M_{\tilde{l}} = 90$ GeV is used, 
the two-body decay of the
$\sim 95$ GeV $\chi^+_1$ or $\chi^0_2$
into a lepton-slepton pair
is kinematically allowed.
As a result, the contributions of the channels
{\boldmath{$e \ell M\!J$}} and {\boldmath{$\nu \ell M\!J$}} 
are considerably enhanced. 
In the case shown in Fig.~\ref{fig:lim_combine_zino},
the channel {\boldmath{$\nu M\!J$}} remains dominant
even in the light sleptons case because the $\chi^+_1$ mass is here
$\sim 80$ GeV.
Only squark decays into the heavier $\chi^0_2$
lead to the enhancement of the channel
{\boldmath{$\nu \ell M\!J$}} shown in Fig.~\ref{fig:lim_combine_zino}c.

Despite the fact that the relative contributions of the
various channels are strongly model dependent,
the upper limits on the Yukawa coupling do not
depend significantly on the scenario considered because the
sensitivity of our analysis is similar in all
gauge channels, and because the covered branching ratio 
is always close to $100 \%$.

\begin{figure}[b]
  \begin{center}
  \hspace*{-1.5cm}\begin{tabular}{cc}
   \mbox{\epsfxsize=0.6\textwidth
      \epsffile{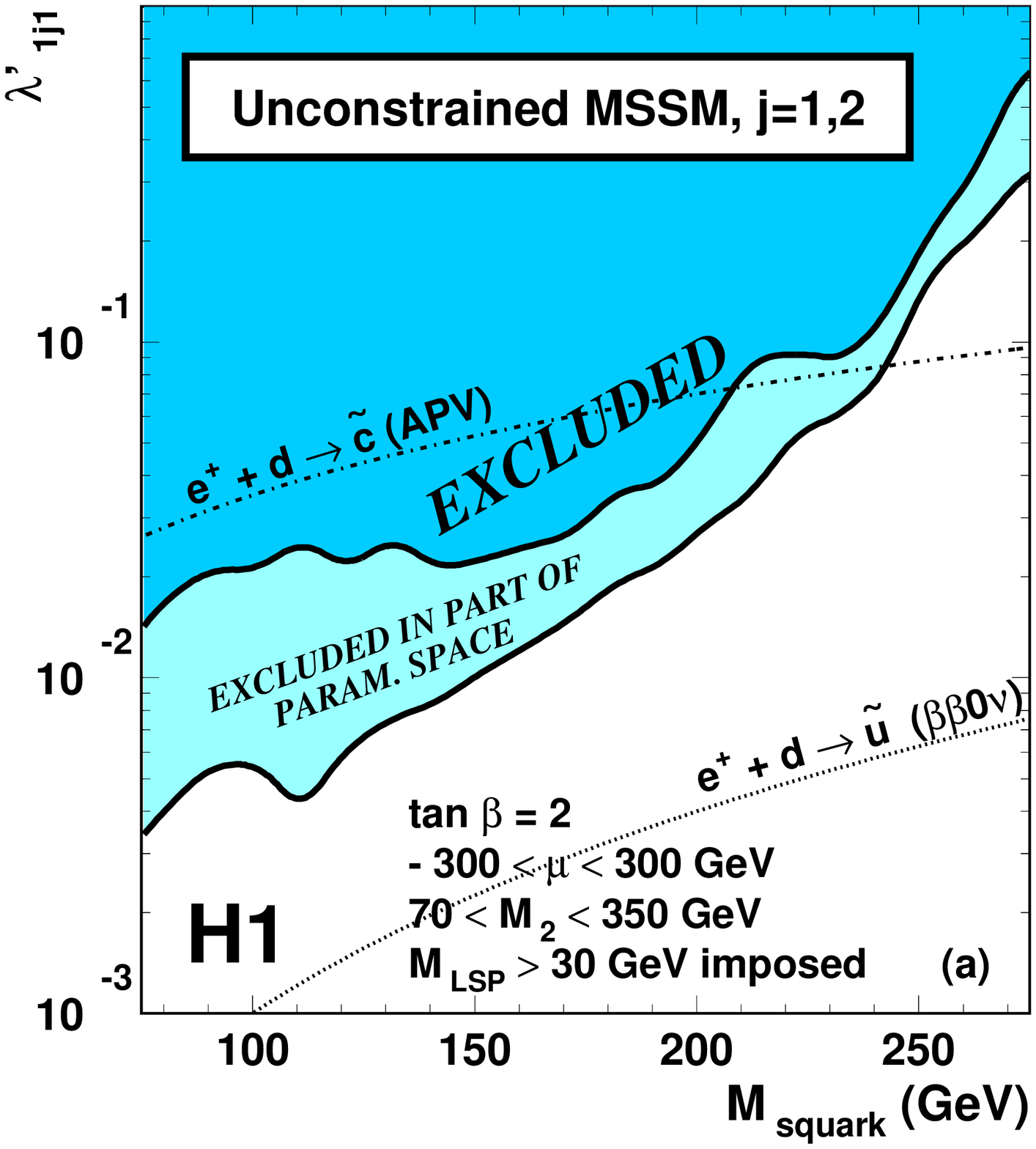}}
   &
   \hspace*{-1.3cm}\mbox{\epsfxsize=0.6\textwidth
      \epsffile{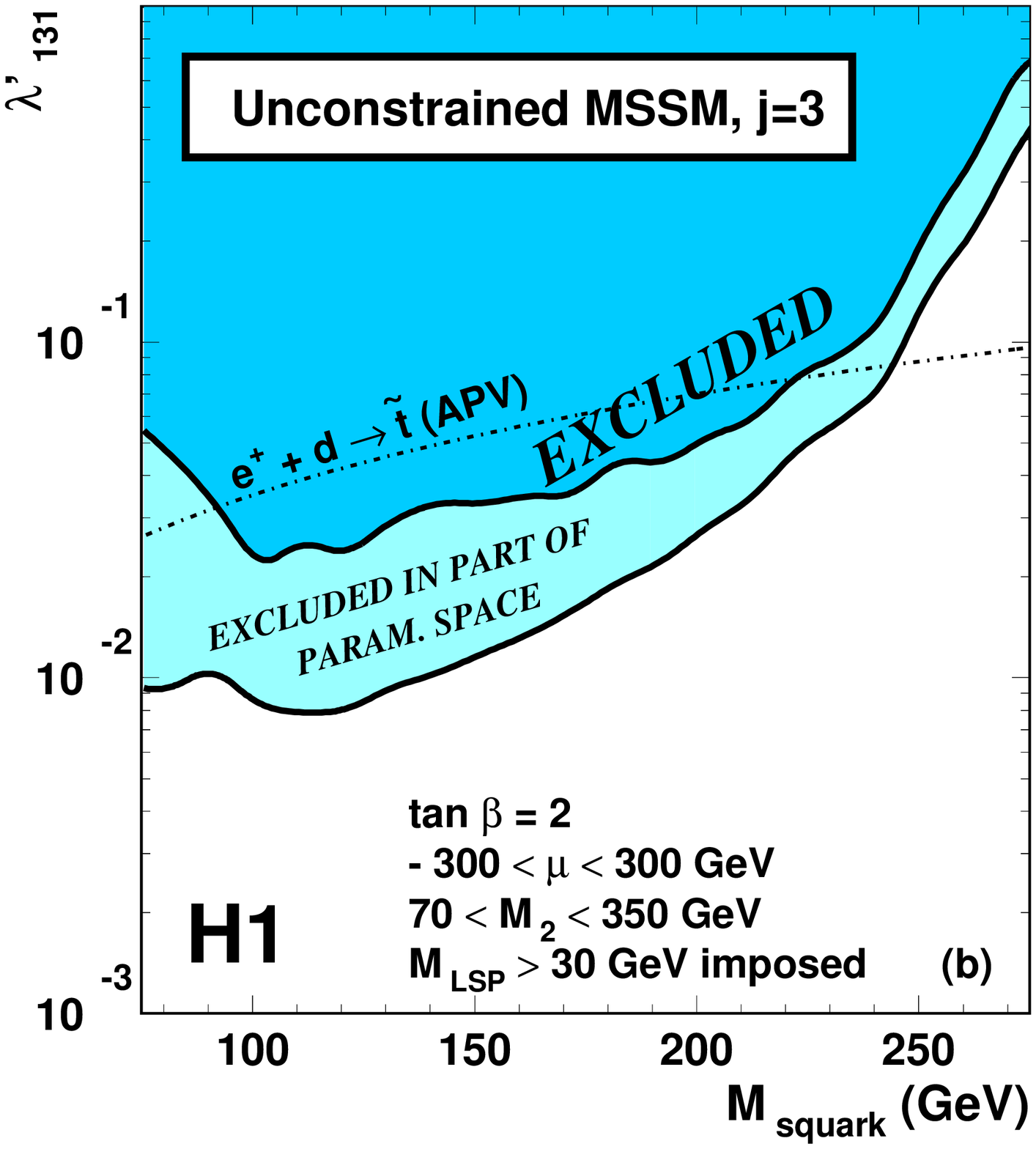}}
   \end{tabular}
  \end{center}
 \caption[]{ \label{fig:scan2}
 {\small  Upper limits at $95 \%$ CL on the coupling
       (a) $\lambda'_{1j1}$ (j=1,2) and
       (b) $\lambda'_{131}$ as a function of the squark mass 
       for $\tan \beta = 2$, in the
       ``phenomenological'' MSSM.
       For each squark mass, a scan of the MSSM parameters 
       $M_2$ and $\mu$ has been
       performed and the largest (lowest) value for the coupling limit
       is shown by the upper (lower) full curve.  
       The dotted curve in (a) indicates the indirect
       bound on $\lambda'_{111}$ from neutrinoless double beta decay
       assuming a gluino mass of $1$~TeV. The dashed-dotted curves
       show the indirect bounds from
       atomic parity violation. }}
\end{figure}

In order to investigate more systematically how the sensitivity
depends on the MSSM parameters,
a scan of the parameters $M_2$ and $\mu$
is performed, for $\tan \beta = 2$. 
The effect of varying the parameter $\tan \beta$
will be studied in the next section.
The mass of the sleptons is set to $90 \GeV$,
the parameters $M_2$ and $\mu$ are varied in the range
{\mbox{$70 \GeV < M_2 < 300 \GeV$}} and 
{\mbox{$-300 \GeV < \mu < 300 \GeV$}}.
Points which lead to a scalar LSP or to LSP masses below
$30 \GeV$ are not considered. This latter restriction, as well as
the lower value for $M_2$, are motivated by the exclusion
domains resulting from $\chi$ searches in \Rp\ SUSY at LEP.
For each point in this $(\mu, M_2)$ plane
the upper bound $\lambda'_{lim}$ on the coupling $\lambda'_{1j1}$
is obtained.
The results are shown in Fig.\ref{fig:scan2}a for 
$\lambda'_{1j1}$ ($j=1,2$) and in
Fig.\ref{fig:scan2}b for $\lambda'_{131}$.
The two full curves in
Fig.\ref{fig:scan2} indicate the maximal and minimal values
obtained for $\lambda'_{lim}$ within the parameter space
investigated.

The spread between these extrema for $\lambda'_{lim}$ 
is small for squark masses above $150 \GeV$
and decreases with increasing squark mass.
Comparing Fig.\ref{fig:scan2}a and
Fig.\ref{fig:scan2}b, the constraints 
on $\lambda'_{131}$ and on $\lambda'_{1j1}$ ($j=1,2$)
are seen to be quite similar.
Only for small squark masses is the sensitivity on
$\lambda'_{131}$ reduced because of the small efficiency
in the {\boldmath{$\nu M\!J$}} channel. 

For a Yukawa coupling of electromagnetic strength 
($\lambda^{'2}_{1j1} / 4 \pi = \alpha_{em}$, i.e.\ $\lambda'_{1j1} = 0.3$)
squark masses below $\sim 260$ GeV
are excluded at $95\%$ CL . 
This extends beyond the mass domain excluded from
relevant searches for scalar leptoquarks performed by
the D0~\cite{D01GENE} and CDF~\cite{CDF1GENE} experiments, which
rule out $\tilde{u}^j_L$ squark masses
below $205 \GeV$ if the branching ratio of the squark into
$eq$ is greater than $\simeq 50 \%$, and below $\sim 110 \GeV$
when this branching is $\sim 10 \%$.
Since in \Rp\ SUSY such a branching can be naturally
small (as seen above)
such leptoquark-like constraints are not very stringent.
Direct squark searches at LEP exclude masses below $\sim 90$~GeV.
Bounds from searches for \Rp\ SUSY carried out at the Tevatron 
do not apply in the unconstrained model considered here.

Our results are also compared in Fig.~\ref{fig:scan2} to 
indirect limits~\cite{INDIRDREINER}.
The production of a $\tilde{u}$ squark via a $\lambda'_{111}$ coupling
is very severely constrained by the non-observation of neutrinoless
double beta decay~\cite{BETA0NU}.
The best indirect limit on the coupling $\lambda'_{121}$
($\lambda'_{131}$),
which could allow for the production of
squarks $\tilde{c}$ ($\tilde t$),
comes from atomic parity violation (APV) 
measurements~\cite{INDIRDREINER,APV}.
For squark masses below $\sim 240$ GeV the H1 direct limits
significantly improve this indirect constraint on
$\lambda'_{121}$ by a factor up to $\sim 3$.
In the case of a non-vanishing $\lambda'_{131}$ coupling,
our results are more stringent than the APV constraints for stop masses
between $\sim 100$ GeV and $\sim 220$ GeV.

The limits on the couplings
$\lambda'_{1j1}$ can be translated into
upper bounds on the couplings $\lambda'_{1j2}$, which
could allow for the production of a resonant 
$\tilde{u}^j_L$ via a fusion between the positron and
a strange quark coming from the proton.
To obtain the limit on $\lambda'_{1j2}$ the ratio of the
$s$ and $d$ quark densities and its uncertainty were taken from a LO QCD 
fit similar to~\cite{Fabfit}, in which various neutrino DIS data 
providing constraints on $s(x)$ were considered.
Resulting upper bounds at $M_{\tilde{q}}=200 \GeV$  
are shown in table~\ref{tab:lim1j2}. The limits were conservatively 
derived taking into
account the $2\sigma$ uncertainties of the parton densities
as given by the fit. 
Using the central prediction for $s(x)$ as given by the MRST
parametrisation, limits on $\lambda'_{1j2}$ do not change by
more than $20\%$.
Table~\ref{tab:lim1j2} also shows 
existing indirect bounds~\cite{INDIRDREINER}
for comparison
and recalls
the bounds obtained on $\lambda'_{1j1}$.
The sensitivity of our analysis to the coupling
$\lambda'_{132}$ is significantly better than that coming
from the leptonic decay width of the $Z$ boson.
No attempt was made to derive limits on couplings
$\lambda'_{1j3}$ due to the large uncertainties on the
$b$ quark density at such high $x$ and $Q^2$.

\begin{table}[hbt]
 \begin{center}
   \begin{tabular}{l|c|c|c|c|c|c}
  & $\lambda'_{111}$   & $\lambda'_{121}$ & $\lambda'_{131}$
  & $\lambda'_{112}$   & $\lambda'_{122}$ & $\lambda'_{132}$ \\
  \hline 
  $\lambda'_{lim}$ (H1)   & 0.05        &   0.05          &   0.05  
  &  0.29             &   0.29          &   0.29 \\
  \hline
  $\lambda'_{lim}$ (indir.)
   & 0.004   &   0.07      & 0.07     &  0.04   & 0.08   & 0.66 \\
   & $\beta \beta 0 \nu$  & APV    & APV    & CC univ.
   & $\nu_e$ mass    & $\Gamma (Z \rightarrow ll)$ \\

 \end{tabular}
  \caption[]
          {\small \label{tab:lim1j2}
               $95 \%$ CL upper bounds on the couplings
               $\lambda'_{1j1}$ and $\lambda'_{1j2}$  for a squark mass of
               $200 \GeV$. Also shown  are
               the indirect bounds obtained from
               neutrinoless double beta decay, atomic parity violation,
               charged current universality,
               the upper bound on the neutrino mass,
               and the leptonic decay width of the $Z$ boson.}
   \end{center}
\end{table}

\subsection{Limits on {\boldmath $\lambda'_{1j1}$} in the Constrained MSSM}

In this section we consider a ``constrained'' 
(supergravity inspired) version of the MSSM
where the number of free parameters is reduced by assuming,
in addition to the GUT relation mentioned in section~\ref{sec:limideri}
between $M_1$, $M_2$
and $M_3$, a universal mass parameter $m_0$ for all sfermions
at the GUT scale. 
The evolution of the sfermions masses towards low scales is 
given by the Renormalisation Group Equations (RGE) and depends on
the gauge quantum numbers of the sfermions. As a result, the sfermions
masses at the electroweak scale are related to each other and to the
parameters determining the gaugino sector.
The model is thus completely determined by e.g.\
$m_0$, $M_2$, $\mu$ and $\tan \beta$ ($m_A$ is set to $300 \GeV$
and we assume no mixing between the
sfermions at the electroweak scale).

For a given value of the $\tilde{u}^j_L$ squark mass, the requirement
of sfermion unification at large scale imposes an upper bound on 
the parameter $M_2$, which is obtained using approximate
solutions for the RGE\footnote{
 The possible effect of the Yukawa couplings
 $\lambda'_{1jk}$ on the RGE has not been taken into
 account here. }.
The upper bound on $M_2$ increases with the squark mass
and is smaller for the stop than for
other squarks.
As before sets of parameters leading to a scalar LSP or to a LSP mass below
$30 \GeV$ are not considered.
This lower bound on the LSP mass forbids too small values
of $M_2$ and hence imposes a lower bound
on the $\tilde{u}^j_L$ mass, which is more stringent in 
case of the stop.
A scan of $\mu$, $\tan \beta$ and $M_2$ is performed
within $-300 \GeV < \mu < 300 \GeV$, $2 \le \tan \beta \le 40$
and within the $M_2$ range allowed by $\tan \beta$ and the $\tilde{u}^j_L$ mass.

\begin{figure}[bht]
  \begin{center}

   \mbox{\epsfxsize=0.6\textwidth
      \epsffile{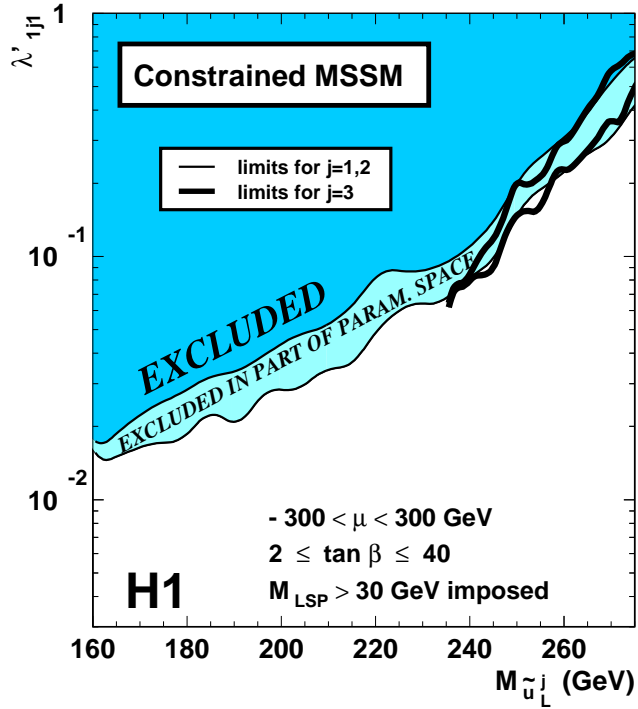}}
  \end{center}
 \caption[]{ \label{fig:scan3}
 {\small  Upper limits at $95 \%$ CL for the coupling
       $\lambda'_{1j1}$ as a function of the mass of the
       $\tilde{u}^j_L$,
       in the constrained MSSM.
       For each squark mass, a scan of the MSSM parameters 
       $\mu$, $M_2$ and $\tan \beta$ has been
       performed and the largest (smallest) value for the coupling limit
       is shown by the upper (lower) curve.  
       The resulting band is contained within the thin curves
       for $\lambda'_{1j1}$ ($j=1,2$) and within the thick ones
       for $\lambda'_{131}$. 
       The requirement on the LSP mass imposes the $\tilde{u}^j_L$
       to be heavier than $\sim 160$ GeV ($\sim 235$ GeV)
       for $j=1,2$ ($j=3$). }}
\end{figure}

The curves in
Fig.\ref{fig:scan3} indicate the maximal and minimal values
for $\lambda'_{lim}$ as a function of the $\tilde{u}^j_L$
mass. 
%
%
%
%
The spread of the domain spanned by the limits
$\lambda'_{lim}$ is quite small, i.e.\ the sensitivity of our
analysis on $\lambda'$ does not depend strongly on the free parameters
of the model, in particular on $\tan \beta$. 
The most stringent limits are usually obtained for
intermediate $\tan \beta$ and are in general
better than those derived previously in the ``unconstrained''
MSSM because
in this range the
sneutrinos can be very light, leading to an enhancement of the
quasi background-free channels {\boldmath{$e \ell M\!J$}} and
{\boldmath{$\nu \ell M\!J$}} via e.g.\ $\chi^+_1 \rightarrow l^+ \tilde{\nu}$.

For a Yukawa coupling of electromagnetic strength, squark masses
up to  values of $260-270 \GeV$
can be ruled out at $95 \%$ CL in the framework
of the constrained MSSM.
Moreover, for a coupling strength $100$ times smaller than
$\alpha_{em}$, the most conservative bound on the mass of the $\tilde{u}^j_L$
obtained from the present analysis still reaches $182 \GeV$.

Searches for \Rp\ SUSY performed at LEP~\cite{L3RPV} also set limits on
the model considered here.
At LEP the mass domain explored by direct searches for squarks 
with \Rp\ couplings is limited by the beam energy. 
However searches for neutralinos and charginos lead to a lower
bound on $M_2$ which, using the RGE's, can be translated into a lower bound
of $\sim 240$ GeV on the squark mass, thus reducing the
allowed mass domain probed in Fig.\ref{fig:scan3} where only
the less stringent condition $M_{LSP} > 30$~GeV was imposed.
The combined searches for $\chi$'s and sleptons at LEP
moreover increase the lower $M_2$ bound for low values of the
$m_0$ parameter. The resulting lower bound on the mass of
first and second generation squarks is close to the current
HERA centre of mass energy.

\subsection{Constraints in the Minimal Supergravity Model}
\label{sec:msugra}

The model considered above can be further constrained
by imposing 
a common SUSY soft-breaking mass term for all scalar fields, and
by assuming that the breaking of the electroweak symmetry is
driven by radiative corrections.
These additional assumptions lead to the so-called {\it minimal}
supergravity
(mSUGRA) model~\cite{MSUGRA}. 
By requiring Radiative Electroweak Symmetry
Breaking (REWSB), 
the modulus of $\mu$ is related to the other model parameters.
The program SUSPECT~1.2~\cite{SUSPECT} is used to obtain 
the REWSB solution for $|\mu|$ when the other parameters are fixed.

Assuming a fixed value for the $\Rp$ coupling $\lambda'_{1j1}$,
our searches can be expressed in terms of constraints on
the mSUGRA parameters, for example on $(m_0, m_{1/2})$ when 
$\tan \beta$, the common trilinear coupling at the GUT scale
$A_0$, and the sign of $\mu$ are fixed.
The parameter $A_0$ enters only marginally in the interpretation
of the results and $A_0$ is set to zero.  
Values of the parameters leading to a LSP lighter than $30 \GeV$
have not been excluded here. However a vanishing efficiency
has been assumed for squarks undergoing a gauge decay ending
by a $\chi$ or $\tilde{g}$ lighter than $30 \GeV$, since 
the parametrisation of the efficiencies (see section~\ref{sec:dismc})
is not valid in this domain.

\begin{figure}[htb]
  \begin{center}
   \begin{tabular}{cc}
   \mbox{\epsfxsize=0.5\textwidth
      \epsffile{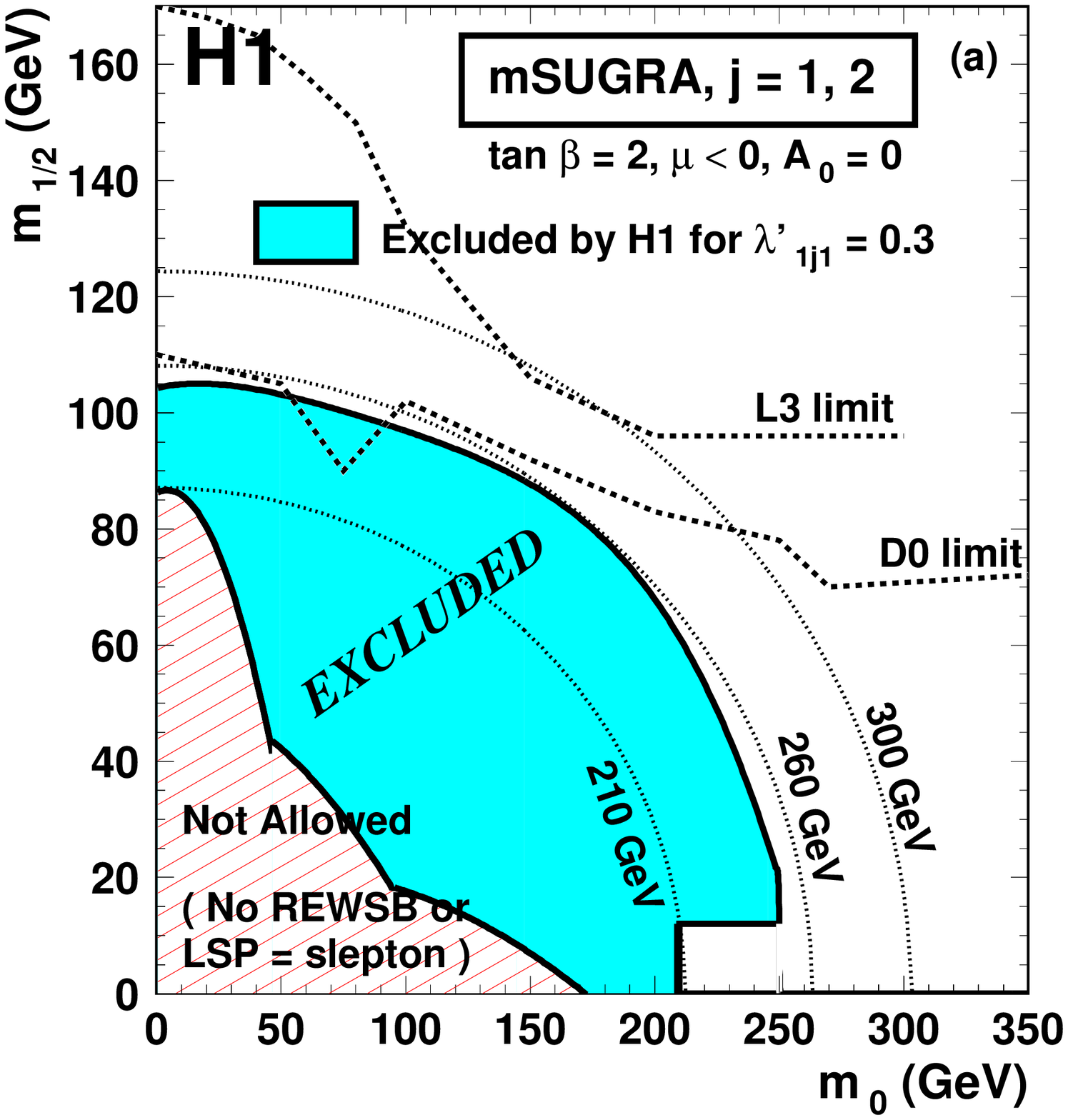}}
   &
   \mbox{\epsfxsize=0.5\textwidth
      \epsffile{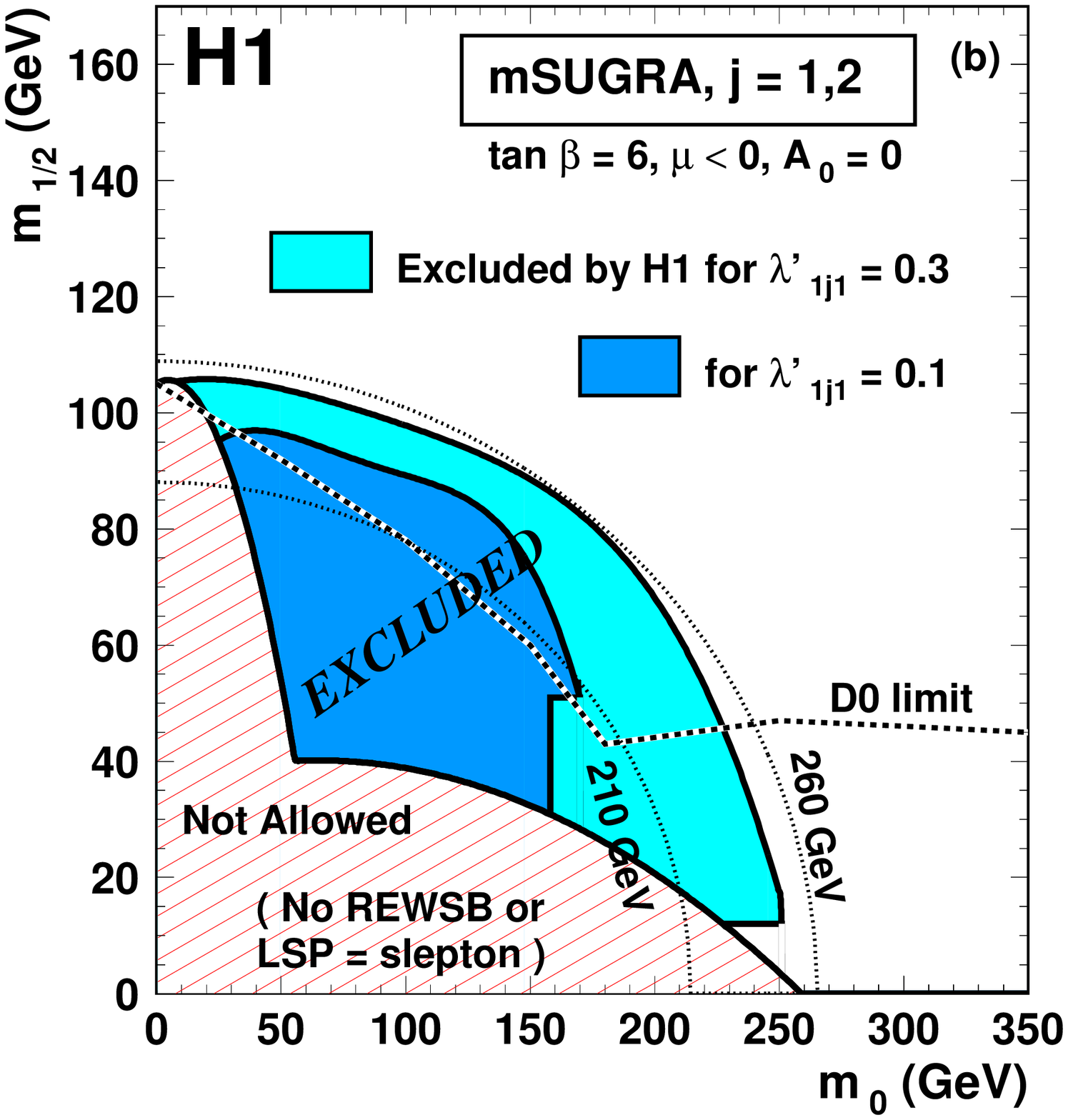}}
  \end{tabular}
  \end{center}
 \caption[]{ \label{fig:sugrat2}
 {\small  Domain of the plane $(m_0, m_{1/2})$ excluded by this
       analysis for $\mu < 0$, $A_0 = 0$ and (a) $\tan \beta = 2$ 
       or (b) $\tan \beta = 6$, for a \Rp\ coupling
       $\lambda'_{1j1} = 0.3$ ($j=1,2$) (light shaded areas).
       In (b) the exclusion domain obtained for
       $\lambda'_{1j1} = 0.1$ ($j=1,2$) is also shown as the dark
       grey area.
       The hatched domains correspond to values of the parameters
       where no REWSB is possible or where the LSP is a 
       sfermion. The regions below the dashed curves are
       excluded by the D0 experiment;
       in (a) also the L3 bound is shown as the dashed-dotted curve;
       these limits do not depend on the value of
       the \Rp\ coupling.
       Two isolines for the mass of the
       $\tilde{u}^j_L$ are also shown 
       as dotted curves. }}
\end{figure}

For $\tan \beta = 2$ and $\mu < 0$, results obtained for a
Yukawa coupling $\lambda'_{1j1} = 0.3$ 
($j=1,2$) are shown in the $(m_0, m_{1/2})$ plane
in Fig.~\ref{fig:sugrat2}a.
%
%
The parameter space where {\mbox{$M_{\tilde{u}^j_L} < 260 \GeV$}}
is nearly fully excluded.
At low $m_{1/2}$ values where the lightest $\chi$'s and the $\tilde{g}$ are
lighter than $30 \GeV$, the sensitivity on the $\tilde{u}^j_L$ mass
however decreases
since the efficiency is conservatively set to zero for
all channels but {\boldmath{$LQe$}}, and reaches $\sim 210 \GeV$.
Fig.~\ref{fig:sugrat2}a also shows
the domain excluded by the D0 experiment~\cite{NIRMALYA}
from searches for SUSY where $R_p$ is violated by a
$\lambda'_{1jk}$ coupling, relying on di-electron events.
H1 and Tevatron results are quite similar at low $m_0$.
However, the mSUGRA parameter space is still more constrained
by the combined searches for $\chi$'s and sleptons carried out at
LEP as shown in Fig.~\ref{fig:sugrat2}a.
LEP and Tevatron bounds do not
depend on the value of the Yukawa coupling.

Results for $\tan \beta = 6$ and $\lambda'_{1j1}=0.3$ or 0.1
($j=1,2$) 
are shown in Fig.~\ref{fig:sugrat2}b.
The excluded domains extend considerably beyond
the region ruled out by the D0 experiment.
This is due to the fact that,
for large values of $\tan \beta$, the lightest neutralino is dominated
by its zino component, so that its decay into $e^{\pm}$
is suppressed. As a result the sensitivity of the 
di-electron D0 analysis
is decreased, while the dominant squark
decay mode is still observable in the H1 analysis via the 
{\boldmath{$\nu M\!J$}} and {\boldmath{$\nu \ell M\!J$}} channels.
LEP limits in~\cite{L3RPV} have not been given for this
value of $\tan \beta$ but the corresponding bounds on
$m_{1/2}$ are expected to be similar to those
shown in Fig.~\ref{fig:sugrat2}a within $\sim 10$ GeV.

We now consider a non-vanishing coupling $\lambda'_{131}$
which could lead to the production of a stop. 
The weak interaction eigenstates $\tilde{t}_L$ and $\tilde{t}_R$ 
mix in this case through
an angle $\theta_t$ to form two mass eigenstates,
labelled $\tilde{t}_1$ and $\tilde{t}_2$ ($\tilde{t}_1$ being
the lightest by convention):
$$ \left(  \begin{array}{c}
   \tilde t _{1} \\
   \tilde t _{2}
\end{array}   \right) = \left(  \begin{array}{clcr}
\cos\theta_{t}    &  \sin\theta_{t} \\
-\sin\theta_{t}    &  \cos\theta_{t}
\end{array}   \right)
\left(  \begin{array}{c}
   \tilde t _{L} \\
   \tilde t _{R}
\end{array}   \right)$$
The production cross-section of the $\tilde{t}_1$
($\tilde{t}_2$) scales as
$\lambda^{'2}_{131}  \cos^2\theta_t$
($\lambda^{'2}_{131}  \sin^2\theta_t)$
since only $\tilde{t}_L$ enters in the $L_1 Q_3 \bar{D}_1$
operator.
Hence, the lightest stop $\tilde{t}_1$ does not necessarily
have the largest production cross-section.
Thus both $\tilde{t}_1$ and $\tilde{t}_2$ are searched for in the analysis.

For channels {\boldmath{$e^-\!M\!J$}}, 
{\boldmath{$e \ell M\!J$}} and {\boldmath{$\nu \ell M\!J$}}
where the signal is integrated over the whole mass range
the fraction of the visible signal  
in a given channel, $k$, is
$\sum_{i=1,2} \beta_{k,i}  \varepsilon_{k,i}  \sigma_i / \sigma_{tot}$,
where $\beta_{k,i}$ is the branching ratio for 
$\tilde{t}_i$ to decay into this channel $k$, 
$\varepsilon_{k,i}$ the corresponding selection efficiency, 
$\sigma_i$ the production cross-section of $\tilde{t}_i$,
and $\sigma_{tot} = \sigma_1 + \sigma_2$ the total signal
cross-section.

For the channels {\boldmath{$LQe$}}, {\boldmath{$eM\!J$}} and {\boldmath{$\nu M\!J$}}
where the signal is integrated over a 
``sliding mass bin"
only the contribution of the state $\tilde{t}_i$ for which the sensitivity
is maximal (i.e.\ which maximises
$\sigma_i  (\sum_k \beta_{k,i}  \varepsilon_{k,i} ))$
is taken into account in the above summation.
The numbers of observed and expected events 
are then integrated
in the mass bin corresponding to $\tilde{t}_i$ only.

\begin{figure}[h]
  \begin{center}
   \begin{tabular}{cc}
   \mbox{\epsfxsize=0.5\textwidth
       \epsffile{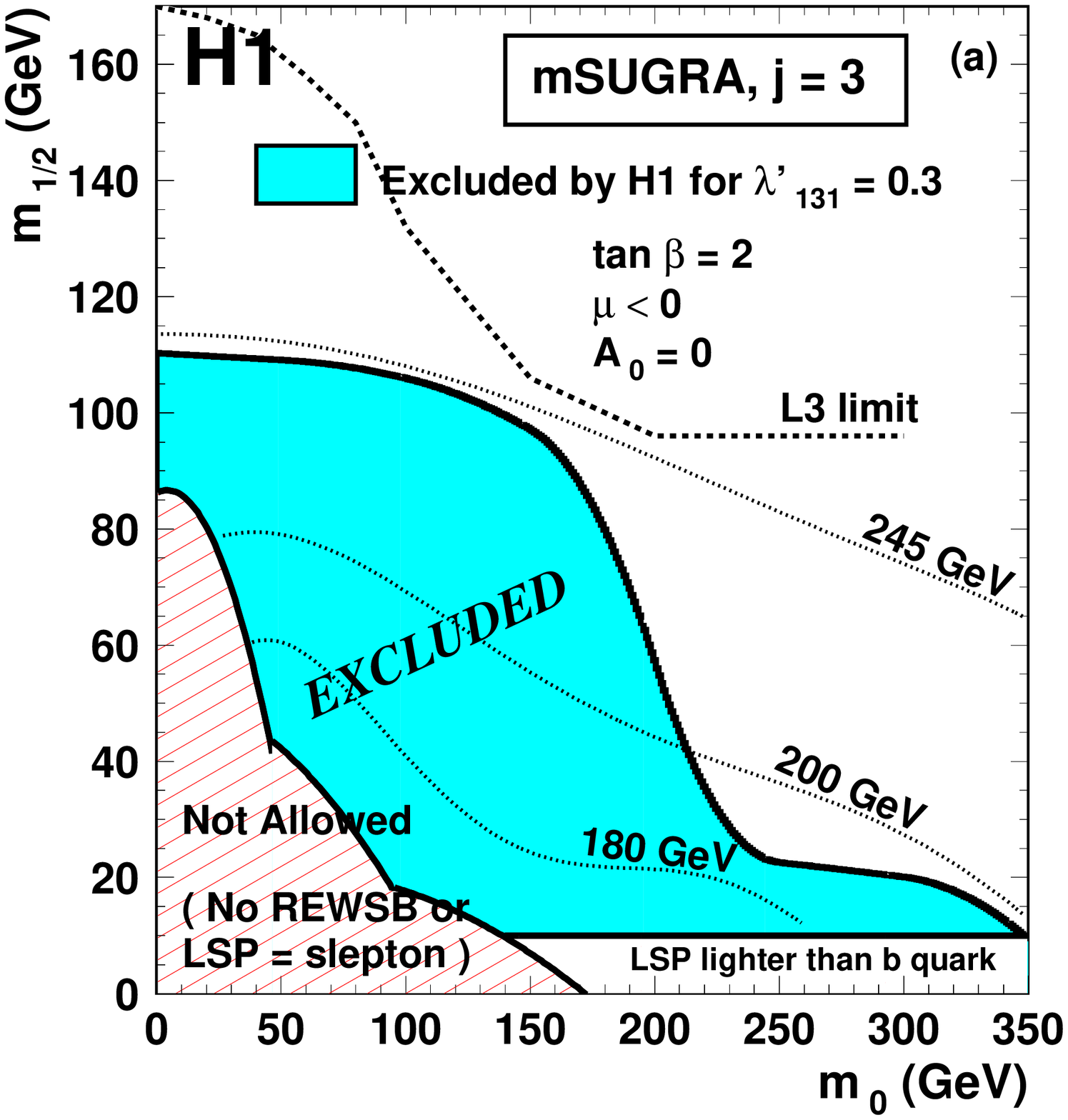}}
 &
   \mbox{\epsfxsize=0.5\textwidth
       \epsffile{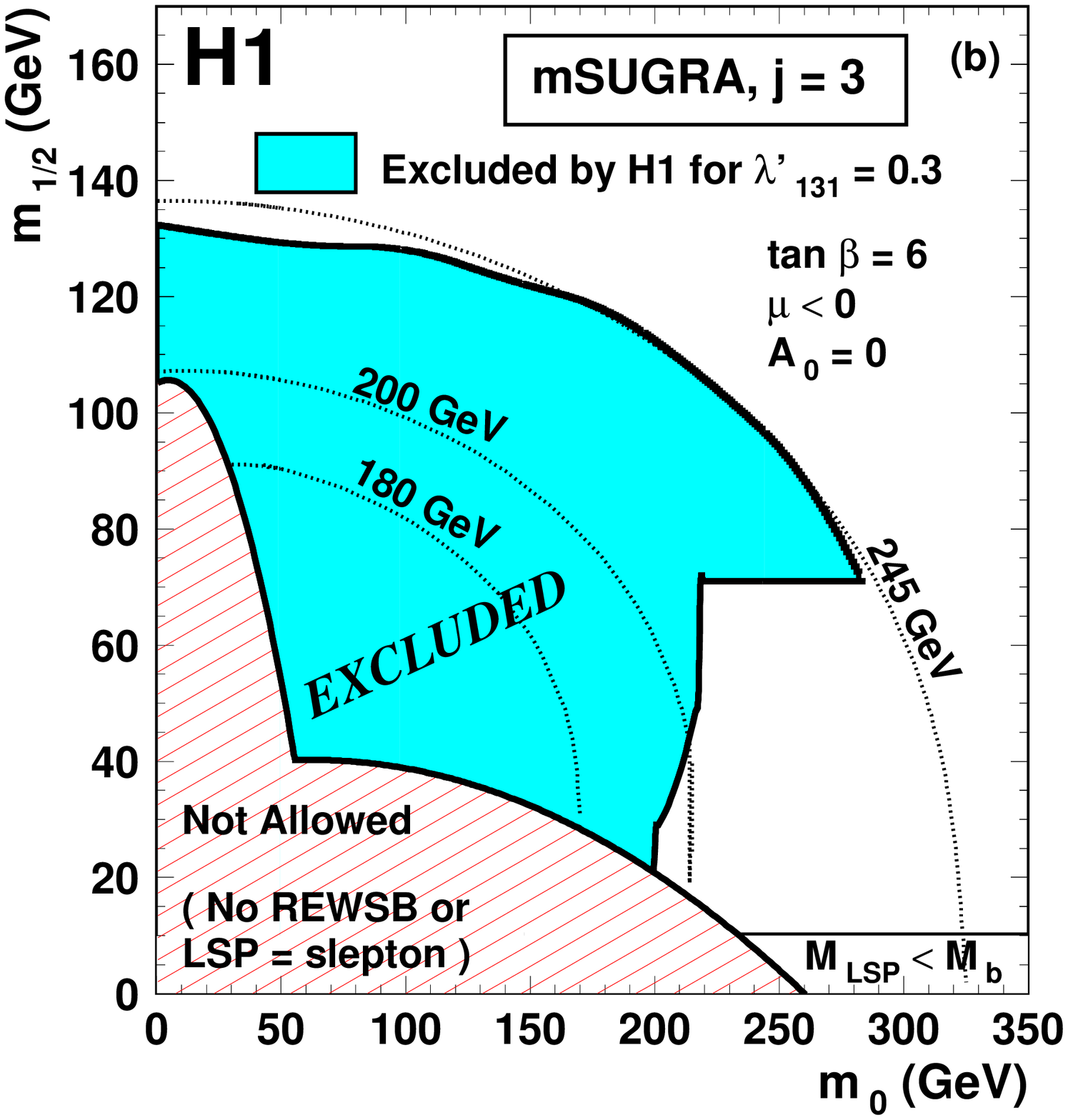}}
  \end{tabular}
  \end{center}
         \caption[]{ \label{fig:sugraj3}
         {\small
       Domain of the plane $(m_0, m_{1/2})$ excluded 
       for a \Rp\ coupling
       $\lambda'_{131} = 0.3$, for $\mu < 0$, $A_0 = 0$ and 
       (a) $\tan \beta = 2$ or (b) $\tan \beta = 6$.
       The hatched domains on the left 
       correspond to values of the parameters
       where no REWSB is possible or where the LSP is a
       sfermion. 
       For too small $m_{1/2}$ (the domain below
       the line $m_{1/2} \sim 10$ GeV), the LSP becomes
       lighter than the $b$ quark and thus is stable
       for a non-vanishing $\lambda'_{131}$ coupling. 
       The region below the dashed curve in (a) is excluded by
       the L3 experiment; this limit does not depend on the
       value of the \Rp\ coupling.
       Three isolines for the mass of the
       $\tilde{t}_1$ are shown 
       as dotted curves. }}
\end{figure}
%

The results are shown in Fig.~\ref{fig:sugraj3} for
$\lambda'_{131}=0.3$, $A_0=0$, $\mu < 0$ and $\tan \beta = 2$ or
$\tan \beta = 6$.
The domain below the line $m_{1/2} \lsim 10$ GeV
is not considered since it corresponds to
cases where 
the only allowed LSP decay into $\nu b \bar{d}$ is
kinematically forbidden.
For $\tan \beta = 2$, the excluded domain
is slightly larger than that ruled out
previously for $\lambda'_{1j1} = 0.3$ ($j=1,2$),
due to the mixing in the stop sector
which leads to $\tilde{t}_1$ masses smaller than
the masses of the other $\tilde{u}^j_L$ squarks.
In particular, larger values of $m_0$ can be probed. 
This remains the case for $\tan \beta = 6$ as long as
$m_{1/2}$ is large enough to ensure that the mass of the lightest
neutralino is above $30 \GeV$. 
When the $\chi^0_1$ becomes too light, the efficiencies
for the channels involving a $\chi^+_1 \rightarrow \chi^0_1$
decay are set to zero, and
the sensitivity to the signal is only provided
by the {\boldmath{$LQe$}} channel or by the decays
$\tilde{t} \rightarrow b \chi^+_1$ followed by a \Rp\ decay of
the chargino. 
As a result, only lighter stops can be probed, for which the 
visible cross-section is large enough.
Note that for both values of $\tan \beta$,
masses of $\tilde{t}_1$ up to $245 \GeV$ can be excluded
for $\lambda'_{131} = 0.3$.
This is slightly smaller than the maximal sensitivity of $\sim 260$ GeV
reached, for the same coupling value, on the 
$\tilde{u}^j_L$ ($j=1,2$) mass (Fig.~\ref{fig:sugrat2}),
or on the $\tilde{t}_L$ mass in the constrained MSSM
when the stop mixing is neglected
(Fig.~\ref{fig:scan3}). This is due to the $\cos^2 \theta_t$
reduction of the $\tilde{t}_1$ cross-section.

For intermediate values of $m_0$,
the L3 sensitivity is slightly better than the
limit obtained from this analysis for $\tan \beta =2$.
In the same part of the parameter space, the H1 limit is comparable with or
slightly extends beyond the expected LEP sensitivity for
$\tan \beta = 6$.

%
\section{Conclusions}
 
We have searched for squarks in $e^+ p$ collisions at HERA
in $R$-parity violating SUSY models. 
No evidence for the resonant production of 
such particles was found in the various channels
considered which cover almost all decay modes.
Mass dependent limits on $R$-parity violating 
couplings were derived.
The limits were set within the unconstrained MSSM, the constrained
MSSM and the minimal Supergravity model.
The model dependence of the results was studied in detail
by performing a scan of the MSSM parameters and was found
to be small.

In the large part of the MSSM parameter space covered by the scan,
the existence of squarks coupling to an $e^+  d$ pair with masses
up to $260 \GeV$ is excluded at $95 \%$ confidence level for
a strength of the Yukawa coupling equal to $\alpha_{em}$.
For a $100$ times smaller coupling strength squark masses
below $182 \GeV$ are ruled out.
This improves the indirect bounds set by low-energy experiments
and, in SUSY models where the sfermion and the gaugino sectors are not
related, extends beyond the reach of other collider experiments.
In models where the sfermion masses depend on
the parameters which determine the supersymmetric gauge sector,
the limits extend beyond the constraints obtained at the
Tevatron collider for intermediate values of $\tan \beta$
and for a Yukawa coupling of electromagnetic strength,
and are comparable with LEP bounds in part of the
parameter space.

\section*{Acknowledgements}
We wish to thank the HERA machine group as well as the H1 engineers and
technicians who constructed and maintained the detector for their
outstanding efforts.
We thank the funding agencies for their financial support.
We wish to thank the DESY directorate for the support
and hospitality extended to the non-DESY members of the collaboration.
We thank the members of the French ``Groupement
de Recherche en Supersym\'etrie" for useful discussions, for
their help in adapting the SUSYGEN generator to $ep$ collisions
and for providing the SUSPECT program.
 
 
{\Large\normalsize}

\end{document}

%% file: h1auts_new.tex
\noindent
C.~Adloff$^{33}$,              
V.~Andreev$^{24}$,             
B.~Andrieu$^{27}$,             
T.~Anthonis$^{4}$,             
V.~Arkadov$^{35}$,             
A.~Astvatsatourov$^{35}$,      
I.~Ayyaz$^{28}$,               
A.~Babaev$^{23}$,              
J.~B\"ahr$^{35}$,              
P.~Baranov$^{24}$
% ...

\bigskip{\it \noindent
% ...
$ ^{4}$ Inter-University Institute for High Energies ULB-VUB, Brussels;
          Universitaire Instelling Antwerpen, Wilrijk; Belgium$^{ c}$ \\
% ...
$ ^{23}$ Institute for Theoretical and Experimental Physics,
          Moscow, Russia \\
 $ ^{24}$ Lebedev Physical Institute, Moscow, Russia$^{ e,h}$ \\
% ...
$ ^{27}$ LPNHE, Ecole Polytechnique, IN2P3-CNRS, Palaiseau, France \\
 $ ^{28}$ LPNHE, Universit\'{e}s Paris VI and VII, IN2P3-CNRS,
          Paris, France \\
% ...
$ ^{33}$ Fachbereich Physik, Bergische Universit\"at Gesamthochschule
          Wuppertal, Wuppertal, Germany$^{ a}$ \\
% ...
$ ^{35}$ DESY, Zeuthen, Germany$^{ a}$ \\
% ...
$ ^a$ Supported by the Bundesministerium f\"ur Bildung, Wissenschaft,
      Forschung und Technologie, FRG,
      under contract numbers 7AC17P, 7AC47P, 7DO55P, 7HH17I, 7HH27P,
      7HD17P, 7HD27P, 7KI17I, 6MP17I and 7WT87P \\
% ...
$ ^c$ Supported by FNRS-NFWO, IISN-IIKW \\
% ...
$ ^e$ Supported by the Deutsche Forschungsgemeinschaft \\
% ...
$ ^h$ Supported by Russian Foundation for Basic Researc
      grant no. 96-02-00019 \\
% ...
}